# An Integration of policy and reputation based trust mechanisms

**Muhammad Yasir Siddiqui**
**Alam Gir**

School of Computing
Blekinge Institute of Technology
SE – 371 79 Karlskrona
Sweden

This thesis is submitted to the School of Computing at Blekinge Institute of Technology in partial fulfillment of the requirements for the degree of Master of Science in Computer Science. The thesis is equivalent to 20 weeks of full time studies.

**Contact Information:**
Authors:
Muhammad Yasir Siddiqui
Address: 3A Lgh 0156 Karlskrona
E-mail: yasirsidiqi@yahoo.com

Alam Gir
Address: 3A Lgh 0156 Karlskrona
E-mail: alamgir.bth@gmail.com

University advisor:
Jenny Lundberg, PhD
jenny.lundberg@bth.se
School of Computing

School of Computing
Blekinge Institute of Technology
SE – 371 79 Karlskrona
Sweden

Internet : www.bth.se/com
Phone : +46 455 38 50 00
Fax : +46 455 38 50 57

# ABSTRACT

**Context:** Due to popularization of internet and e-commerce, more and more people getting involved in online shopping market. A large number of companies have been transferred to the internet where online customers have been increased due to easy access. The online business facilitates people to communicate without knowing each other. The e-commerce systems are the combination of commerce behavior and internet technologies. Therefore, trust aspects are positive elements in buyer-seller transactions and a potential source of competitive e-commerce industry.

There are two different approaches to handle the trust. The first approach has a solid authentication set of rules where decisions are made on some digital or logical rules called *policy based trust mechanism*. The second approach is a decentralized trust approach where reputation assembled and shared in distributed environment called *reputation based trust mechanism*.

**Objectives:** In this thesis, the strengths and weaknesses of policy and reputation based trust mechanisms have been identified through systematic literature review and industrial interviews. Furthermore, the process of integrated trust mechanism has been proposed.

**Methods:** The integrated trust mechanism is proposed through mapping process, weakness of one mechanism with the strength of other. The proposed integrated trust mechanism was validated by conducting experiment with buyer/seller scenario in auction system.

**Conclusion:** The analysis of collected results indicated that proposed integrated trust mechanism improved the trust of buyer against eBay and Tradera. At the end, we have discussed some key points that may affect trust relationship between seller and buyer. Furthermore, there is a need for further validation of proposed trust mechanism in auction system/e-commerce industry.

# CONTENTS

# 1 INTRODUCTION

In this chapter, background of the selected research area, problem domain, aims and objectives of the thesis are described. Further in this chapter, research questions and selected research methodologies for this thesis study is presented.

## 1.1 Background

Now a day's internet has become a business hub because of its increased usage among people. E-commerce industry and online customers have been increased rapidly due to easy access. Customers to customer (C2C) e-commerce such as auction systems are more popular between individual internet users. C2C, auction systems has simple transaction process which makes this type of online shopping more popular among others. E-commerce applications are growing and getting more complex. Volume of e-commerce trading increased three times from the volume of 2007 that prevent potential users to have trust in newly arrived sellers/buyer in e-commerce industry [1].

Rapid increase of e-commerce especially for auction systems users facing more problems about trust, make it hard for new sellers and buyers to establish trustworthy relationship [1]. The current auction systems can be web applications or stand-alone software. Auction system provides ability for users to post their products for bidding. In most cases both buyer and seller don't know about each other while making a deal of transaction. From buyer aspect it's hard for buyer to trust on new seller for establishing trustworthy business partnership.

Web of trust is an important area in both industry and academia. Many trust mechanisms have been developed so far, each has a different approach and characteristics about trust. Trust layer in semantic web refers to trust mechanisms which involve verification process that the source of information refers who the source claims to be and how much trustworthy it is. Verification process involves encryption and signature mechanisms that allow any consumer of this particular information to verify the source of the information. Reputation and authentication were focused according to the work previously done by different researchers on trust mechanisms [2].

Marsh was the first one who analyzed Trust as a computational concept in the distributed artificial intelligence domain [3]. Computational concepts are used currently over the web as rating systems which clearly describe positive ratings for particular web content in a particular environment. G. Zacharia proposed a model considering buyer's credit in the calculation of seller's credit, which is believed to make the evaluation more reliable. Relationship between consumer and seller while they have done transaction is the main consideration in evaluation of trust in G. Zacheria model [4]. The results of simulation indicate that it had improved effects based on G. Zacharia model in the situations like reputation collision or reputation slander [5]. Furthermore, Tale l.0 Abdessalem described trust models and mechanisms for calculating trust between pair of users. Under his research, he explained how each participant is responsible for their ratings from other participants in distributed environment.

Many social networks, e-commerce and web content systems are using rating systems such as Smart Information Systems, Smart Assistants *"Based on Semantic-Web-data and is using ontology information to map customer needs to technical product attributes"* [6]. Smart Information system is providing an easy way to locate data within the web trustfully.

Usually we can define reputation as the trust amount inspired by a particular person in a specific domain of interest [3]. Reputation evaluated according to its expected economic outcomes is regarded as asset creation in *"Trust in a Cryptographic Economy"* [7]. Another similar study was conducted by Heski Bar-Isaac on seller reputation where he introduced a framework which embeds a number of different approaches to find the seller reputation [8]. Recommended trust evaluation model is proposed by Tianhui You and Lu Li for e-commerce applications based on trust evaluation model considering the consumers purchasing preference in e-commerce industry [9]. Tianhui model can simulate the results that indicate it had better effects, confronted with fraud behavior and trust of buyer in seller. In all the studies described above, the main focus of researchers was policy and reputation based mechanisms of trust [3][4][5][6][7][8][9].

Policy based trust mechanism has a solid authentication set of rules such as trusted certification authorities and signed certificates. Policy based trust mechanism consists on binary decisions. These decisions can be made on pre defined policies, in response resources/services may be allowed or denied. Second trust mechanism is a reputation based which involves "soft computations" i.e. rating systems. Many rating systems are more popular over the web which are based on these reputation based trust mechanisms. Reputation based mechanisms has been more useful in semantic web or Peer-to-Peer i.e. auction systems in e-commerce industry [10]. Both policy and reputation based trust mechanisms are addressing the same problem, to establish trust between interacting parties in distributed and decentralized environment but from different perspectives and have different type of settings to act upon. Trust management will be more benefit from an intelligent integration of both policy and reputation based trust mechanisms. In some situations, trust can be better achieved from policy, while in other situations benefits may be attained by the use of reputation in such an integrated approach. An integrated mechanism will enhance the existing trust management tools and can be very effective [10].

## 1.2 Purpose

The purpose of this thesis work is to propose and implement an integrated mechanism of both policy and reputation based trust mechanisms. The proposed process is based on the identified strengths and weaknesses of the two commonly used trust mechanisms i.e. policy and reputation based trust mechanisms. Furthermore, the part of integrated mechanism is implemented as a prototype on auction system. The experiment conducted to validate the effectiveness of integrated trust mechanism by comparing it to both policy and reputation based trust mechanisms. The comparatives study was conducted with proposed trust mechanism with eBay and Tradera.

## 1.3 Problem Domain

In e-commerce industry, auction systems needs more trust for establishing trustworthy relationship between two parties. Usually new sellers don't have any ratings which represents reputation based trust mechanism in such auction system environment [11]. Buyers have less trust in new sellers because new seller doesn't have any reputation on a platform [12]. Seller reputation plays an important role to increase the trust of buyer in seller because buyers often choose sellers with respect to their reputation [13]. Basically there is a need to suggest such a mechanism which can help to build trust on new sellers for auction systems. Few platforms offer newcomers to pay entry fee in order to consider trustworthy, which could be an alternative approach [14]. In online marketplace this approach would be applicable but not very popular in buyer and seller relationship. To our best knowledge, there is lack of mechanism that can build trust between two parties. So we are encouraged to suggest an integrated trust mechanism of both policy and reputation based trust mechanisms. This could be helpful in auction systems to improve trust on new sellers.

## 1.4 Aims and Objectives

The aim of this thesis work is to implement and validate an integrated trust mechanism of both policy and reputation based trust mechanisms. On the basis of strengths associated with both policy and reputation based trust mechanisms, an integrated trust mechanism may implement in such a way that can address main issues of both trust mechanisms. Furthermore, experiment has been conducted to validate the effectiveness of the implemented integrated trust mechanism by comparing with both policy and reputation based trust mechanisms in industry.

The major objectives of this thesis study are:

- Identifying the strength of policy and reputation based trust mechanisms
- Identifying the weaknesses of policy and reputation based trust mechanisms
- Implementation of an integrated trust mechanism
- Validation of integrated trust mechanism through experiment

# 1.5 Research Questions

Three research questions are proposed which depict the reason for conducting this research.

*RQ1. What kinds of circumstances are more suitable for policy respective reputation based trust mechanisms in auction systems?*

The answer to this question highlights the strengths and weaknesses of both mechanisms in different circumstances.

*RQ2. How to integrate both reputation and policy based mechanism to increase chances of trust?*

On the basis of strengths associated with both policy and reputation based trust mechanisms, an integrated trust mechanism is defined in such a way that can address the main issues of both trust mechanisms.

*RQ3. Could there be benefits of using both reputation and policy based trust mechanisms in establishment of new seller relation with customers in auction systems?*

Experiment is conducted to validate the effectiveness of implanted integrated trust mechanism. Formulation of hypothesis was used to verify the correctness of collected data from experiment. Statistical and hypothesis testing was done to answer the question i.e. trust level of customer increased using proposed integrated trust mechanism against eBay and Tradera. Detailed discussion is presented after analysis of collected data.

# 1.6 Research Methodology

Creswell defines research as a study that goes beyond the influences of personal ideas and experiences of an individual. A researcher's work is primarily based on the utilization of some research methods and techniques [15]. Creswell describes three types of methods used for research i.e. Qualitative, Quantitative and Mixed research.

In this thesis, we are following both qualitative and quantitative approaches. Each question is answered with proper selected research method. Two different Qualitative methods were used for data collection in order to answer RQ1 and RQ2. Systematic literature review used to identify the strength and weaknesses of both policy and reputation based mechanisms. Systematic literature review leads us towards better understanding of concepts/characteristics about both trust mechanisms in e-commerce industry. Interviews were conducted from industrial experts in order to verify our findings from literature review and to avoid researcher's biasness. The results of literature review and industrial interviews summarized to answer the RQ1 and RQ2. Data collected through industrial interviews/systematic review helped to propose and design an integrated trust mechanism. Quantitative approach was used in order to answer RQ3, where an experiment was conducted to validate the proposed integrated trust mechanism.

Table 1 Research questions and their respective methodologies

| Research Questions | Methodology |
|---|---|
| Research Question 1 | Systematic Literature Review/ Interviews |
| Research Question 2 | Systematic Literature Review/ Interviews |
| Research Question 3 | Experiment/Results |

## 1.7 Research Design

The graphical representation of stages involved in study process are described in figure 1.0

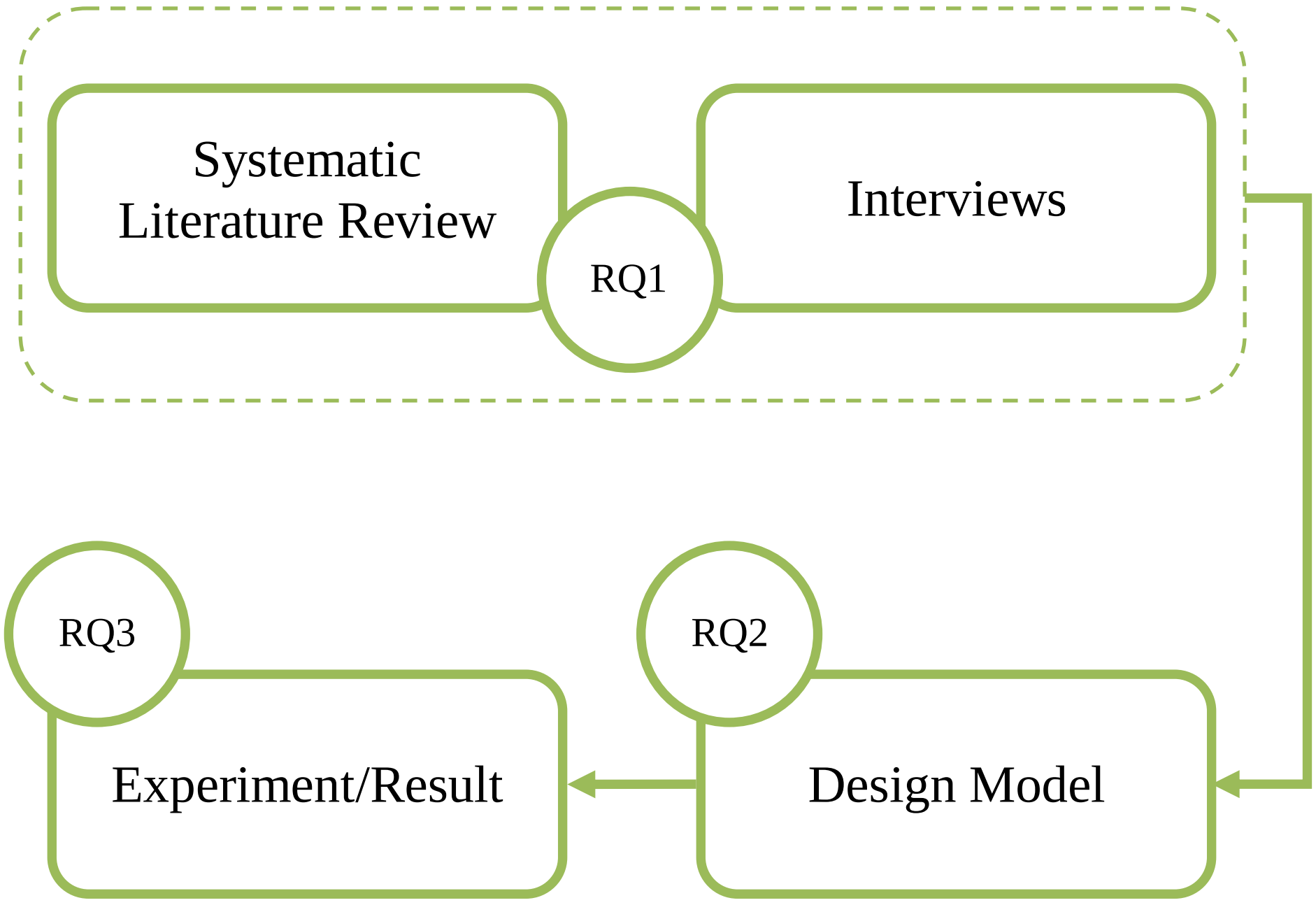

Figure 1 Research Design

The validation of integrated trust mechanism was done through an experiment. The experiment was designed based on data collected through systematic literature review along with industrial interviews. In the first step, strengths and weaknesses of both trust mechanisms were identified then on the behalf of those identified strengths and weaknesses a mapping process was applied. The strength of one mechanism was mapped to weakness of other mechanism, mapping process gives us clear idea how we can integrate both trust mechanisms. The design model was assisted by mapping process to resolve the identified weaknesses in integration. Furthermore, to validate and support the integration process an experiment was conducted. The RQ3 was answered through the results of experiment.

## 1.8 Thesis Structure

**Chapter 1** (Introduction): This chapter describes the background, problem domain, purpose of study, research aims, objectives and adopted research methodologies for this thesis study.

**Chapter 2** (Background: policy and reputation based trust mechanisms): In this chapter, background work and basic ideas related to both trust mechanisms are presented.

**Chapter 3** (Strength and weaknesses of policy and reputation based mechanisms): In this chapter, systematic literature review and industrial interviews are presented to identify the strength and weaknesses of both policy and reputation based trust mechanisms and to identify the benefits of an integrated mechanism.

**Chapter 4** (Process of integrated trust mechanism): In this chapter, on the basis of strength and weaknesses of both policy and reputation based trust mechanism, an integrated trust mechanism is proposed.

**Chapter 5** (Experiment and results): In this chapter, details are given about experiment. The experiment design and variables were used to conduct the experiment, in order to validate the designed trust mechanism. Collected data from experiment used to discuss the effectiveness of proposed integrated trust mechanism.

**Chapter 6** (Conclusion and future work): In this chapter, conclusion and future work is presented.

**References**

**Appendix**

# 2 BACKGROUND: POLICY AND REPUTATION BASED TRUST MECHANISMS

Usually policy and reputation mechanisms are used in different organizations for trust establishment in the industry. Both policy and reputation based trust mechanisms have been used in different environments and have different set of rules to act upon. Policy based trust mechanism is a *centralized approach* where binary trust decisions has been made on some digital and logical rules. Reputation based trust mechanism as a *decentralized approach* where trust decision has been made on the basis of personal experience and experience of other entities i.e. rating/feedback. In some cases trust may not be fully achieved either through policy or reputation based trust mechanism. Industry may get benefits from an intelligent integration of both policy and reputation based trust mechanisms. The purpose of this thesis work is to overcome weaknesses of policy and reputation based trust mechanisms by introducing an integration of both trust mechanisms.

In this chapter, work previously done on both policy and reputation based trust mechanisms are described in details to give basic understanding of both trust mechanisms.

## 2.1 Policy Based Trust Mechanism

Policy based trust mechanism has a solid authentication set of rules such as trusted certification authorities and signed certificates. Policy based trust mechanism consists of binary decisions. These decisions can be made on the basis of given credentials by an entity, in response resources/services should be allowed or denied [10]. In the following section, polices used to establish trust are summarized.

### 2.1.1 Network Security Credentials

The applications are performed on the basis of given credentials by an entity, where credentials are some set of information regard to trust. Different polices used a broad set of information as credentials to make trust decisions. A common example of a credential e.g. signing in to any online site on the web, a valid user name with a correct password must be given to gain access. According to defined policy, this information proves that the given user is the verified administrator. It is a bidirectional approach for establishing trust, where the user must keep his password secret. Credentials maybe implemented by using security certificates having properties about an entity. Kerberos protocol is used to securely exchange verifiable credentials [16].

### 2.1.2 Trust Negotiation

Trade-off between privacy and earning trust is the focal point in trust negotiation. Winslett and colleagues focused to earn trust in a particular context by revealing specific credentials, where credential privacy is lost after credential revealing [17]. Winslett implement an architecture called TrustBuilder, which provide mechanisms for addressing privacy and earning trust trade-off. Traditional security techniques (e.g., authentication, encryption etc) were used for establishing trust in TrustBuilber. TrustBuilder provides the concept of credential chain, that is if A trust on the credentials of B and B trust on the credentials of C, then A have some trust on the credentials of C. Many different trust negotiation languages (e.g., trust management language RT, PeerTrust and Ponder etc) were designed to exchange credentials and perform efficient search chain [18] [19].

### 2.1.3 Security Policies/Trust Languages

Both security and trust are co-dependent, related concepts for singular purposes. Mostly trust related policy languages designed for use in the semantic web i.e. KAoS, Rei etc related to access control and exchange of credentials [18]. KAoS encouraged the use of same policy in distributed heterogeneous environment while Rei allowing each party to identify their own policy [18]. Recent efforts describe the expression and representation of trust while creating security policies. Nelson work provides a formal policy language where access control is determined by user's level of trust [20]. Some languages e.g., XACML and SAML treat trust and security separately while providing means for authentication and authorization.

### 2.1.4 Distributed Trust Management

Trust management broadly described the problem facing by credentials, as credentials are also subject to trust. Early work on trust management was found in PolicyMaker, which suggest the separation of security and trust. PolicyMaker encouraging individual systems to have their own separate trust polices with respect to global authentication and security system [21]. KeyNote is another system, provides a standard independent policy language and more features with respect to PolicyMaker [22]. Policy language presented in KeyNote is independent from the used programming language. Some researchers defined trust as what to earn after credentials verification and still preferred a hard security approach [23].

### 2.1.5 Credential Type Effects

Trust is measured with respect to given credentials. A credential may be a resume, text chat, id or picture of an entity. It is assumed that type of a credential affects the amount of trust or distrust received, where some type of credential affect more than other in certain scenarios [24].

## 2.2 Reputation Based Trust Mechanism

Personal experience and the experience of other entities in the form of ratings/feedbacks were used to make a trust decision in reputation based trust mechanism. In the following section, reputation based trust mechanisms are described in details.

### 2.2.1 Decentralization and Referral Trust

Reputation is a decentralized trust approach, where individuals are allowed to make trust decisions rather than to rely on a single centralized process [25]. Yu and Singh described a reputation management system, where agents determining trust on the basis of information they receive from other agents. According to Yu and Singh reputation management avoids hard security approaches while they use trust information from external sources, known as referral trust [26]. Sabater focused on information context while presenting their solution to referral trust, that who can be trusted and for which context they can be trusted [27].

### 2.2.2 P2P Networks Trust Management

Reputation based trust applications are commonly used in P2P networks and grids. Anyone is allowed to upload any kind of data with any name on P2P networks. On the basis of P2P uploads, EigenTrust algorithm determine a global reputation value for each entity. Reputation system in P2P network using protocol and algorithms for referral trust management [28]. Abrer and Despotovic used statistical analysis for scalable computation of determining trust reputation [29]. Another example is XRep protocol where feedback history was used to determine the best host by automatic vote.

### 2.2.3 Trust Metrics in Web of Trust

The reputation is a transitive process of trust computation. For example, one might trust on an author or book because of its publisher where the publisher is recommended by one of its friend. In such a transitive process each entity maintains reputation information for other entities, thus creating a web of trust [30]. Trust and reputation information's are expressed through ontologies. The ontologies allow quantification of trust used in algorithms to make trust decision about entities. Trust quantification in algorithms often refers to trust metrics [31]. A simple example of transitive trust is, if A trust B and B trust C, then A trust C. Zhang used a set of hypothesis and experiments considering types of links, type of resources and type of trust in the known entity for transferring trust over the web [32]. The problem of controversial users over the web was presented by Massa. According to Massa the local calculated value of trust will be accurate in contrast to globally computed value (value in the web of trust) [33]. Ding and his colleagues present a method, using both context and referral trust to compute trust over the web [34].

### 2.2.4 Application Specific Reputation

Reputation based systems are used by different specific applications according to their own environments. Ad-hoc networks use their reputation system for selecting node in a network for transferring data. In the ad-hoc networks, nodes can indirectly monitor the performance of other nearby nodes to select trustworthy node for transferring data [35]. Allocating tasks to the best performing agent is another specific application of reputation [36].

# 3 STRENGTHS AND WEAKNESSES OF POLICY AND REPUTATION BASED TRUST MECHANISMS

In this chapter, we present a systematic literature review and industrial interviews conducted from experts (who have minimum two years of experience in semantics), to identify the strengths/weaknesses of both policy and reputation based trust mechanisms, to identify the benefits of using an integrated mechanism. On the basis of the results collected from systematic literature review/industrial interviews, an integrated trust mechanism of both policy and reputation based mechanisms will be proposed.

## 3.1 Systematic Literature Review

The systematic literature review defined by Kitchenham is to identify, evaluate and interpret relevant available research material in order to answer a research topic of interest or research questions [37]. Contribution of individuals in any fashion to systematic literature review considered as primary studies. Systematic literature review considered as secondary study. In this thesis, we will closely follow Barbara Kitchenham guidelines for conducting systematic literature review.

There are three main phases to conduct a systematic literature review [37].

- Planning the Review
- Conducting the Review
- Reporting the Review

The first phase associated with the need of conducting review along with development of review protocol. A review protocol defines the guideline which leads toward the process of systematic literature review.

The second phase associated with the following sub phases.

- Identification of research
- Selection of primary studies

- Study quality assessment
- Data extraction and monitoring
- Data synthesis

The third phase is single stage phase, where results of the systematic literature review are presented.

## 3.1.1 Planning the Review

### 3.1.1.1 Identifying the Need of Systematic Literature Review

The systematic literature review gives us an opportunity to accommodate and summarize the related research which has previously been done. We gathered the related research to find out empirical evidence that focus on strengths and weaknesses of both policy and reputation based trust mechanisms. As our aim is to propose and implement a new integrated trust mechanism of both policy and reputation based trust mechanisms. We assume that latest research will be more fruitful in the process of proposing and implementing an integrated trust mechanism. Furthermore any gap related to the current study is suggested for further investigation.

## 3.1.2 Development of Review Protocol

Review protocol is essential part that describes detailed blue print for conducting systematic literature review. A pre-defined protocol provides a way for selecting primary studies which can trim down the possibility of researcher biasness [37].

The search terms were applied before conducting the systematic literature review to know the previous work done by others. The systematic review in thesis should be based on existing research along with proposed research fills the gap in current body of knowledge [37]. The result of findings before systematic literature review shows that most of the research has been carried out in recent ten years. The selection of research papers/articles is based on years from 2000 to 2010. We were able to search research articles without boundaries even then our aim lead us towards recent research articles, details are in above section. The reason behind the specified time period was to get an overview of recent research carried out on policy and reputation based trust mechanisms. The research which has been carried out in recent years can indicate any gap related to policy and reputation based trust mechanisms.

### 3.1.2.1 Search Strategy

The search strategy consists on selection of research material and online resources based on search strings. Search strings and relevant resources are listed below:

#### 3.1.2.1.1 Search String

The aim for performing systematic literature review was to find out relevant research work that has been done on policy and reputation based trust mechanisms. Preliminary search was carried out to extract relevant studies with the following search strings.

```
(Trust model AND (policy based mechanism OR reputation based
mechanism OR new sellers))
(Trust model AND (policy based mechanism OR reputation based
mechanism OR eBay))
(Trustworthy AND (policy based mechanism OR reputation based
mechanism OR new sellers))
(Trust model AND (policy based benefits OR reputation based
benefits OR eBay))
(Trust model AND (policy based strength OR reputation based
strength OR new sellers))
Policy based mechanism AND reputation based mechanism in auction
system
Policy based mechanism AND reputation based mechanism in C2C
```

After performing the search with the help of these queries, a total of 2497 papers have been found, described in table 10. Furthermore, inclusion exclusion criteria will reduce the number of research articles.

#### 3.1.2.1.2 Recourses Utilized

The online resources utilized for this systematic literature review are as under:

- IEEE Explorer
- ACM Digital Library
- Inspec (www.iee.org/Publish/INSPEC/)
- ISI (Online search engine database)
- EI Compendex (www.engineeringvillage2.com)

### 3.1.2.2 Criteria for Study Selection

In section below, the relevant articles are selected from primary studies. The study selection criteria based on the following inclusion and exclusion criteria.

#### 3.1.2.2.1 Inclusion Criteria for Study Selection

The inclusion and exclusion criterion is defined to select primary research papers and articles. Primary selected articles will reviewed for further most relevant studies and data extraction purpose. The inclusion criterion is used to identify the primary studies related to strengths/weaknesses of both policy and reputation based trust mechanism. Following is detailed inclusion and exclusion criteria which will be applied on selected studies.

1. The research papers or articles are selected that defines the trust mechanisms, any one of policy or reputation or other sort of relevant information.
2. The research papers or articles are selected that may address systematic literature review, case study, surveys, experiments or analysis reports.
3. The research papers or articles are selected that explain any sort of comparative analysis especially strengths and weaknesses related to trust management.
4. The research papers/articles are selected which they have some sort of cross reviewed.
5. The research papers or articles are selected that provide freely available full text.

#### 3.1.2.2.2 Exclusion Criteria for Study Selection

The article(s) that did not match with inclusion criteria as discussed above, were excluded from selection of research papers/articles.

### 3.1.2.3 Procedure for Study Selection

The primary criterion was used to identify the article weather it is relevant to our topic of interest or not. Selection of primary study requires investigating some key points about selected article in inclusion/exclusion criteria.

- Title of the research paper or article
- Abstract of the research paper or article
- Conclusions of the research paper or article

Inclusion of the article is dependent on above mentioned sections; full article reading fortified if the primary reading about the article satisfy the inclusion criteria.

### 3.1.2.4 Study Quality Assessment Check Lists

The quality check list was prepared based on different sections presented in research articles. These sections include introduction, research methodology, process of reports/results conducting and conclusion section. These checklists will be used for the evaluation of research articles selected in primary study.

#### 3.1.2.5 Strategy Used for Data Extraction

The data extraction strategy defines the procedure for extracting knowledge from selected research articles [37]. Data extraction was based on specific and general information described in research articles, more details are given in below sections.

##### 3.1.2.5.1 General Information

The general information of selected research articles was documented, are listed below:

- Title of the selected Article
- Name of Author(s)
- Name of Conference/Journal/ Date of Publish/Presented
- Relevant Search String(s) utilized to retrieve research article
- Database used to retrieve the research article
- Date of Publication

The specific information about selected research article was documented, described in appendix A.

#### 3.1.2.6 Synthesis of Extracted Data

The data synthesis section defines to collect and summarize the results of primary research articles. The collected articles were found distinct from each other based on research methodology and their outcomes. Qualitative synthesis was appropriate to document the results of the relevant research articles with respect to appropriate research questions.

### 3.1.3 Review conducting

Systematic literature review was conducted in following steps.

#### 3.1.3.1 Research identification

Systematic literature review was conducted to find the maximum number of studies as possible relevant to the research questions of this thesis study [37]. The review protocol explicitly defines the search strategy for performing systematic literature review. A general approach is to break down the research questions into small questions and more individual facts [37].

*What kinds of circumstances are more suitable for policy respective reputation based trust mechanisms in auction systems?*

- What are the strengths and weaknesses of both policy and reputation based mechanisms?
- In which environment policy based trust mechanism is more suitable respective reputation based trust mechanism?

*How to integrate both reputation and policy based trust mechanism to increase chances of trust?*

- What factors to consider in the integration of both policy and reputation based trust mechanisms?
- Could an integrated trust mechanism of both policy and reputation based trust mechanisms will be beneficial in case of seller /customer trust relationship?

On the basis of these research questions search strings were defined by using ANDs/ORs operators. An iterative search strategy was adopted where trail searches were conducted for verification of the search strings. The search strategy was explicitly explained in the review protocol section 3.1.2.1, on the basis of search strategy a preliminary search is carried out to identify the relevant literature data from different online and electronic resources. In addition to digital libraries other relevant resources e.g. books, company articles, etc were also consulted to carry out relevant literature data.

#### 3.1.3.2 Primary Studies Selection

Two main steps were performed in the selection of primary study. Title, abstract and conclusion of the articles were studied in the first step for the selection of relevant research studies. In the second step, inclusion and exclusion criteria was applied on the selected studies. Selected conferences articles and books which are relevant to our research topic are given in the table 11 in appendix A.

A total number of 97 articles were scanned in this systematic literature review and 21 were selected. The selected research papers are listed in the table 12 in appendix A.

Search strings defined in the review protocol were used for searching relevant articles, journals and databases. Some non-relevant articles were rejected on the basis of inclusion and exclusion criteria after conducting detailed study. For example, while searching different trust mechanisms, other articles concerning social trust etc were also displayed, which are non-relevant to the current systematic literature review and research topic were ignored.

## 3.1.4 Study Quality Assessment

Quality assessment is performed on the selected primary research articles on the basis of their structure i.e. Introduction section, research methodology, gathered results, conclusion etc details are in section 3.1.2.4. The quality assessment procedure selects primary research article that provides relevant information about the topic of interest.

## 3.1.5 Data Extraction

In this phase, all the extracted data from the primary study were gathered and documented according to specific and general information described in research articles, as mentioned above in the review protocol 3.1.2.5, it is an easy way of extracting relevant information from the selected primary research study. The results of primary research articles were collected and summarized. Qualitative synthesis was used to document the results of the relevant research articles. All the extracted data was cross-checked in order to avoid missing any relevant information.

## 3.1.6 Review Reporting

In this single phase, the results of systematic literature review are presented with respect to research questions. In the following sub-sections the results of systematic literature review are presented.

### 3.1.6.1 Policy Based Trust Mechanism

Policy based trust mechanism mostly used in environments having strict security requirements [62]. It is a bidirectional trust mechanism, exchanging credentials to establish trust from the scratch in a semantic web environment where different parties make interaction initially unknown to each other [38]. This mechanism is commonly used in access control decisions [10]. These decisions has been made on the bases of given credentials provided by unknown entities and a set of trust policies, whether to allow or deny access to a specific service. Different set of rules using by different trust agents defined trust policies, on the basis of which trust decisions has been made. Policy based mechanism is a binary approach for making access control decisions and referred strong and crisp approach as well. Services of a trusted third party may be used for issuing or verification of credentials in policy based mechanism [10]. Languages having well-defined semantic are used for the implementation of policy based trust where decision based *"non-subjective"* attributes certified by certification authorities (e.g., via digital credentials). Policy based trust mechanism is indented for systems having tough security requirements. Policy based trust mechanism is also preferred for systems where the temperament of information used in authorization process or where people performs sensitive transactions e.g., financial and health services [10]. Strength and weaknesses associated with policy based trust mechanism in the literature are presented in the following sections.

##### 3.1.6.1.1 Strength of Policy Based Trust Mechanism

In this section, strengths associated with policy based trust mechanism found from literature are presented.

The policy based trust mechanism is more secure approach for establishing trust is compared to reputation based trust mechanism. Policy based trust mechanism is a binary approach for making trust decisions i.e. an entity will be allowed or deny, decision is dependent on provided credentials [10]. The policy based trust mechanism is increasing trust in case of sensitive transactions (e.g., financial and health services) via internet because of its strong security mechanisms [40]. Some more benefits of policy based trust mechanism are listed below.

- Policy based trust mechanisms are efficient and bidirectional approach in establishing trustworthy relationships [39].
- Improved security and privacy.
- Policy based trust mechanism is well suitable for specifying whom allowed to access a specific recourse/service [40].
- Insuring customer satisfaction by fast and reliable business transactions and thus producing good customer relationship, as customer feels they are really part of the business growth [41].
- Policy based trust mechanism are using all trustworthily relevant information (signature, age, nationality, identity etc) in the form of credentials.
- Provide functionalities, explanations and answering questions, that how certain information have been trustworthy [42].
- Usually policy based mechanisms establish trust directly between two parties instead of involving a third party [38].
- Only trustworthy customers will be allowed to access specific information.
- Policy based trust mechanism promote product and increase market shares [41].
- Strong and crisp approach for establishing trust [40].
- Sometimes a trusted third party services may be used for the verification of certificates.
- Promote long term relationship with business partners [41].

##### 3.1.6.1.2 Weaknesses of Policy Based Trust Mechanism

In this section, weaknesses associated with policy based trust mechanism in the literature are presented.

According to the literature, the implementation of real world polices are more complex [38]. Often irrelevant information is required in pre-registration phase and there is a chance of disclosing private information like credit card number etc in policy based trust mechanism [39]. Some more drawbacks of policy based trust mechanism are listed below.

- Often, the information required in pre-registration phase is not relevant to the services client willing to access [39].
- Most of the times, customers don't show interest to disclose their private information thus they leave application in pre-registration phase [43].
- Difficulties lies due to context based nature of trust the same agent may change their trust depending on policy for different contexts [44].
- Policy based trust mechanisms, based on a set of given credentials which are also subject to trust [43].
- Neither party is willing to reveal their credential before their opponent.
- In policy based trust mechanisms clients doesn't have choices, such mechanisms act upon binary decisions.
- Customers who are willing to disclose their private information have more concern about the security of their private information in such systems.

#### 3.1.6.2 Reputation Based Trust Mechanism

Many online markets are using reputation or feedback for promoting trust in transactions. Different items of online companies are available for bidding over the web at any time. Reputation based systems encouraged buyers and sellers to rate each other positively or negatively after each transaction [45]. Many online sellers like eBay, amazons etc provide feedback option as well for saving valuable text comments of the customers [68]. Net reputation score of the seller's displayed automatically with each item he/she lists on the auction page and thus repute able sellers have chance to earn more profit and their product quality. Buyers can watch these ratings and text comments before start bidding [45].

##### 3.1.6.2.1 Strength of Reputation Based Trust Mechanism

In this section, strengths associated with reputation based trust mechanism in the literature are presented.

The rating/feedback system in reputation based trust mechanism leads people towards decision making process i.e. whom to trust. The reputation can encourage honest and trustworthy sellers while discourage dishonest sellers [46]. The rating/feedback options in reputation based trust mechanism promote long term relationship with business partners. Some more benefits of reputation based trust mechanism are listed below.

- Reputation based trust mechanism is an easy approach for maintaining trust, where a user fills out simple online form or most times with a single mouse click.
- Reputation based trust mechanism avoids private communication between both sellers and buyers, which encourages a buyer to bid high in such a monitory based system [45].
- More experienced sellers, having more feedback will be able to describe their items in a best manner by changing item title, spelling correctly etc [45].
- Probability of sale and price maybe changed with reputation.
- Reputation encourages new bidders entering the auction system [45].
- Good reputation encourages sellers to sell high quality products.
- Encourages seller have low reputation to get a healthier market with low price and provide verity of quality services [46].
- Feedback system allows sellers to be sustained with high quality products and still earning non-negative profit [46].
- Reliability information of individuals transmitting to third parties by the word of mouth, some of whom will be future trading partners [48].
- Reputation systems rely on indirect repository, which someone trusts you because you are trustworthily to others [48].
- Sellers and buyer can rate each other in reputation system that can effect on the current system or in the whole market.
- Discouraging less reliable sellers to join the marketplace.
- Members can rate the feedback of others as well, i.e. how much useful is a member’s feedback in reputation systems [63].

##### 3.1.6.2.2 Weaknesses of Reputation Based Trust Mechanism

In this section, weaknesses associated with reputation based trust mechanism in the literature are presented.

Many reputation systems currently in the market are too positive from seller’s point of view where negative rating/feedback rarely effects seller’s overall reputation [45]. The use of multiple identities is another problem in such mechanisms [47]. Problems related to reputation based trust mechanism in literature are listed below.

- Buyers don’t know for how long a seller is in the market [45].
- A single entity maybe able to establish many identities and can rate same service object multiple times [47].
- Gaining high reputation by providing many low value services.
- When the same service is offered by many different channels, a single entity will be able to rate only the chosen one [47].
- Re-entry to the community with different identity. Many times an entity whether it is a seller or buyer having bad reputation leaves the community and re-enters with a different identity [47].

- New sellers may not be trusted as they have no reputation at all.
- Seller act honestly in the start by providing high quality services over a period of time for gaining high reputation, then provides low quality services to get profit from high reputation [2].
- Providing unfair ratings that do not reflect the authentic opinion of the rater [47].
- Most of the times people don’t provide feedback it all.
- Most of the times positive feedback increased in reputation but negative feedback don’t affect them.
- Lack of assurance in honest reputation, one party maybe blackmail other by providing false negative or positive reputations [67].
- Required explicit and item specific trust ratings.

## 3.2 Industrial Interviews

Interviews can be considered as an effective way of extracting and eliciting relevant research related information by interviewing a domain expert. Interview is a technique which used to collect qualitative data [49]. The reason of conducting interviews may differ that can fulfill multiple objectives. Interview can either be conducted face to face or telephonic/online. In total four semis structured industrial interviews were conducted for this study. Semi structured interviews were conducted because it provides two way communication where open ended questions are asked to get maximum information on the research topic.

### 3.2.1 Purpose of Interviews

Interviews were conducted to know the industrial viewpoints related to both policy and reputation based trust mechanisms. The benefit of conducting industrial interviews is that, it provides detailed information based on the personal experience of individuals with trust mechanisms. The personal experience of professionals may not written in literature, they can validate finding from literature or can give some suggestions upon their experience in industry. Interviews questionnaire were formulated for two main purposes. The first purpose is to formulate the interview questions based on the strength and weaknesses of both policy and reputation based trust mechanisms described in systematic literature review, in order to validate and avoid researcher biasness. The second purpose was to figure out industrial view point about both trust mechanisms and to get their suggestions for possible solutions of problems in each policy and reputations based trust mechanism that will be considered in the process of integration. RQ1 and RQ2 will be answered on the basis of systematic literature review along with industrial interviews.

## 3.2.2 Selection of Interview Subjects

Peoples who involved in trust layer of semantic web are selected for interviews. People who are directly involved in the development and management of auction system were considered in this regard to gather precise and useful information. Interviewees are selected from reputable organizations and all of them have the same classification i.e. who have minimum two years of experience in semantics. Conducting interviews from selected subject gives us broader aspect of the trust mechanisms in the form of multiple perspectives.

## 3.2.3 Study Instruments

Four study instruments were designed to know about the industrial aspects related to strengths/weaknesses of both policy and reputation based trust mechanisms. Most of the questions in these study instruments are formulated based on systematic literature review. These questions were asked to avoid the research biasness about the specific topic, to get the opinions of experts on both trust mechanisms and their suggestions for possible solutions of problems in both policy and reputation based trust mechanisms.

- The design of first instrument is based on benefits of policy based trust mechanism. This study instrument contains most of the strengths of policy based trust mechanism found from literature, in order to validate and know suggestions from industry professionals.
- The design of second instrument is based on weaknesses related to policy based trust mechanism. This instrument leads towards their experience that might have come across during practice.
- The design of third instrument is based on benefits of reputation based trust mechanism. Different questions are formulated to know the points of view from professionals.
- The design of fourth instrument is based on weaknesses of reputation based trust mechanism. The experience/suggestions of professionals are necessary to validate the weaknesses and get some solutions of those weaknesses.

At the end of the questionnaires, other question derived to know the independent point of view from industry professionals that will be consider in process of integration. The questions asked as study instrument were primarily qualitative in nature that can be viewed as interview questionnaire in appendix B.

## 3.2.4 Interviewing

Each interview is conducted online on Skype due to geographical distribution, in duration of 35 to 40 minutes. A short description of research topic was presented before asking questions from the interviewees. Important points were manually written on the paper during interview. The results collected after interviews were transcribed, as the important points separated from general discussion of interview and from the answers of questionnaire. The transcribed form of interviews can be viewed in appendix B.

## 3.2.5 Validity Threats

The most relevant validity threats related with studies are described. The internal and external validity threats associated with systematic literature and industrial interview are discussed below.

### 3.2.5.1 Internal Validity

The threat in systematic literature review associated with researcher biasness is tried to overcome by following quality assessment criteria along with the use of well known databases. Poorly designed interview questionnaires can affect the outcome of research. To overcome this threat interview questionnaire were designed based on issues and problems associated with both trust mechanisms. The questionnaires of interview were relevant to literature study but needs more industrial emphasis. The formulation of questionnaire was done by mutual discussions that can overcome the threat of missing important questions related to both trust mechanisms.

### 3.2.5.2 External Validity

The external validity threats can be minimized by generalizing the outcome of study in different settings on a small scale. [15].

The systematic literature review conducted on study material from 2000 to 2010. The main purpose of selecting recent years was to gather recent research articles on both trust mechanisms. There is a threat that we may missed any relevant weakness or strength published before selected years associated with any trust mechanism. In order to overcome this threat industrial interviews were conducted where independent opinions/suggestions of experts helped us to know current trends in industry. We also give all the interviewees an introduction about the topic of research before conducting the interviews which may enhance their interest in research topic.

## 3.3 Results and Analysis of the Data Collected Through Systematic Literature Review and Interviews

In this section, results and analysis of systematic literature review and industrial interviews are presented. Strengths and weaknesses of both policy and reputation based trust mechanisms are found from literature. Industrial interviews are conducted in order to validate findings from systematic literature review and to figure out possible suggestions/solutions of problems from experts about both trust mechanisms. The suggestions/solutions from industry professionals will be considered in further process of integration. The identified strength and weaknesses of both policy and reputation based trust mechanisms have been classified into further factors on the basis of their similarities i.e., security, feedback, cost, bad image etc.

Summary of the strengths and weaknesses associated with the policy based trust mechanism presented in the following section. Furthermore, the table below shows the factors of identified strengths in policy based trust mechanism on the basis of their similarities.

Table 2 Classification of the strengths in policy based trust mechanism

| Factors | Strengths in policy based trust mechanism |
|---|---|
| Enhanced Security | • Increase trust in sensitive transactions via internet<br>• Bidirectional and efficient approach for establishing trust<br>• Only trustworthy users will be allowed to access specific information<br>• Strong and crisp approach of establishing trust, strong binary decisions i.e. an entity will either be allowed or denied on the basis of his credentials |
| Customer Satisfaction | • Producing good customer relationship by fast and reliable transactions<br>• Provide functionalities, explanations and answering questions, that how certain information have been trustworthy |
| New Entity | • New customer is reliable due to the verification of their credentials<br>• Only trustworthy customers will be encouraged to join the market |
| Avoid Frauds (duplicate ids) | • Access to specific resources will be allowed after verification<br>• All trustworthily relevant information is useful i.e. (id, credit card etc)<br>• Same agent will not be able to re-enter the system with a different identity |

Graphical representation of strength associated with policy based trust mechanism is given below.

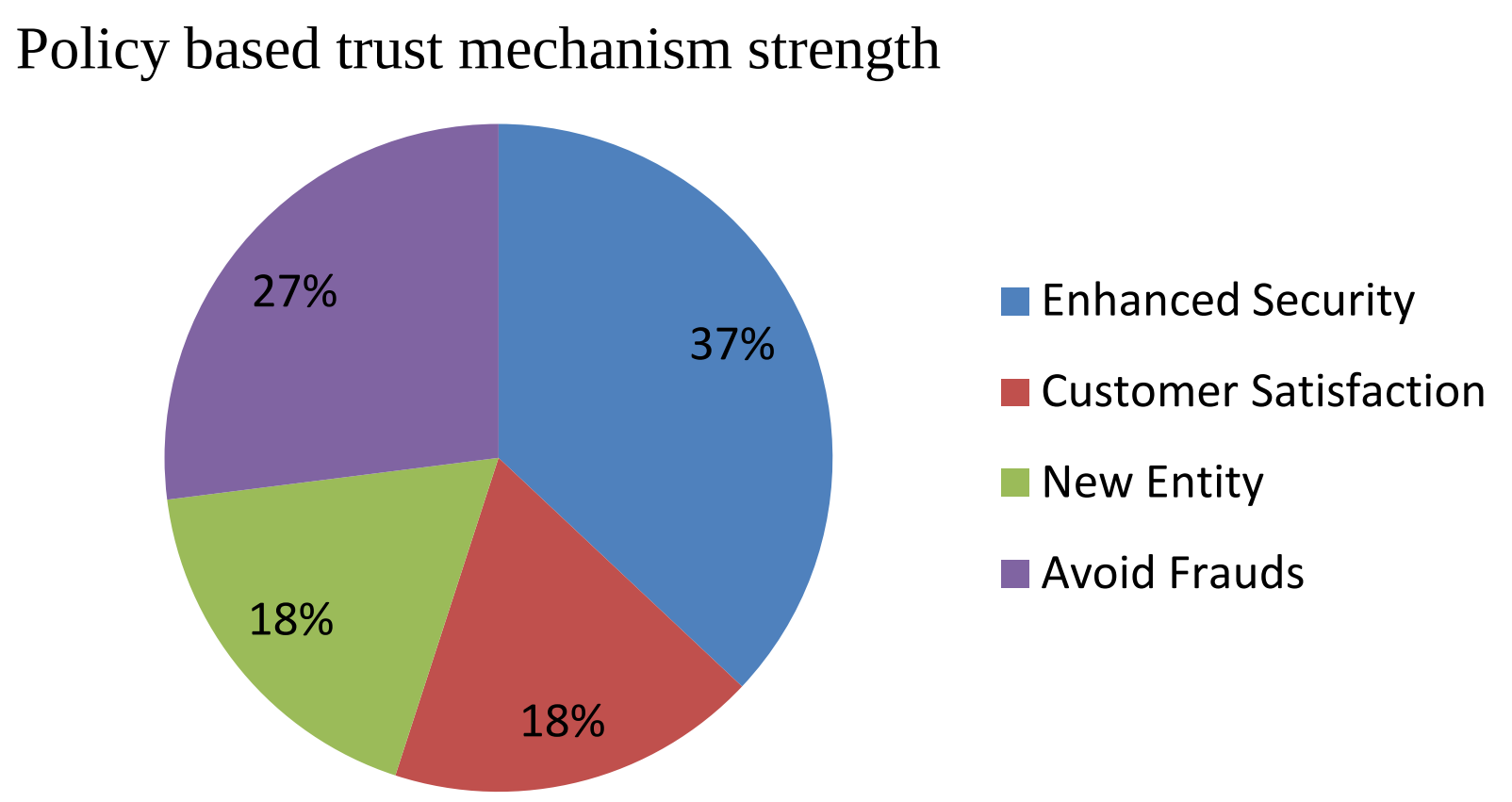

Figure 2 Perceived strength of policy based trust mechanism from the literature and Interviews

Enhanced security is one of the main strength in policy based trust mechanism highlighted by different researchers in systematic literature review along with all the interviewees were agreed as explained in appendix B [10][38][39][44]. Due to reliable transactions policy based trust mechanisms are thought to be satisfactory [39][41][42]. Policy based trust mechanism encourages new seller to start business in e-commerce domain [40][41]. According to the data collected from systematic literature review and interviews the ratio of fraud is less in strong policy based trust mechanisms [10][38][41].

In the following table, the identified weaknesses of policy based trust mechanism have been classified into further factors on the basis of their similarities.

Table 3 Classification of the weaknesses in policy based trust mechanism

| Factors | Weaknesses in policy based trust mechanism |
| --- | --- |
| Time Consuming | • Often information required in pre-registration phase are not relevant to services the client willing to access<br>• Neither party is willing to reveal their credential before their opponent |
| Complex Implementation | • Real world polices are difficult to implement.<br>• Trust expression and representation is complex job while creating security policies<br>• Binary decision system where client don't have choice, they can either be allowed or deny |
| Security Threat | • Credentials are also subject to trust<br>• Disclosing private information of someone, e.g. credit card number etc<br>• The same agent may change their policy for different context |

Graphical representation of weaknesses associated with policy based trust mechanism is given below.

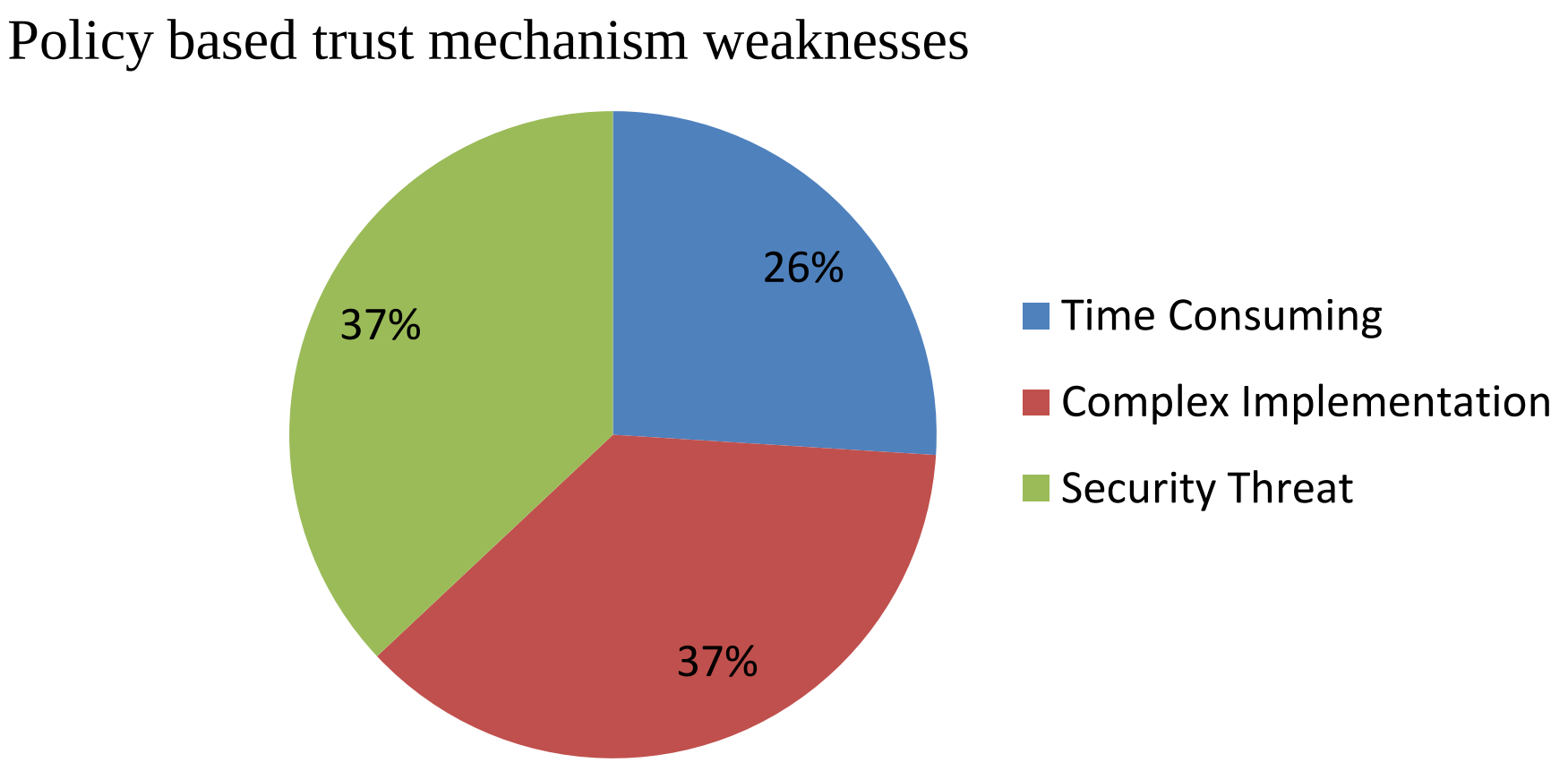

Figure 3 Perceived weaknesses of policy based trust mechanism from the literature and Interviews

Irrelevant information asked in pre-registration phase is time consuming factor. [10][38][43]. Complex implementation is another problem according to literature review and industrial interviews [10][38]. As found from the literature that information provides by entities in the form of credentials are also subject to trust [10][43].

The identified strengths of reputation based trust mechanism have been classified into further factors on the basis of their similarities.

Table 4 Classification of the strengths in reputation based trust mechanism

| Factors | Strength in reputation based trust mechanism |
|---|---|
| Third Party Involvement | • People trust or distrust on a specific entity or item because of its rating<br>• Feedback provide valuable information to buyers about sellers<br>• Reputation avoids from private communication between seller and buyer |
| Increasing Sale Ratio | • Feedbacks helps sellers to describe their items in best manner by changing item title, picture etc<br>• An item price maybe changed due to its rating or feedback information received from different entities<br>• Good reputation encouraged sellers to sell high quality products |
| Increase Business Partners | • Rely on indirect repository that someone trusts you because you are trustworthy to others<br>• Reliable information of individuals transmitting to third parties by the word of mouth, some of whom will be future trading partners<br>• Encouraging new bidders to join the system |
| Feedback | • Buyers are encouraged to leave valuable comments after each transaction<br>• A specific feedback can also be rated in such a reputation system<br>• Provides valuable information about each item displaced by sellers |
| Implementation | • Reputation system can be easily implemented<br>• Flexible and simple approach of maintaining trust |

Graphical representation of strength associated with reputation based trust mechanism is given below:

Reputation based trust mechanism strength

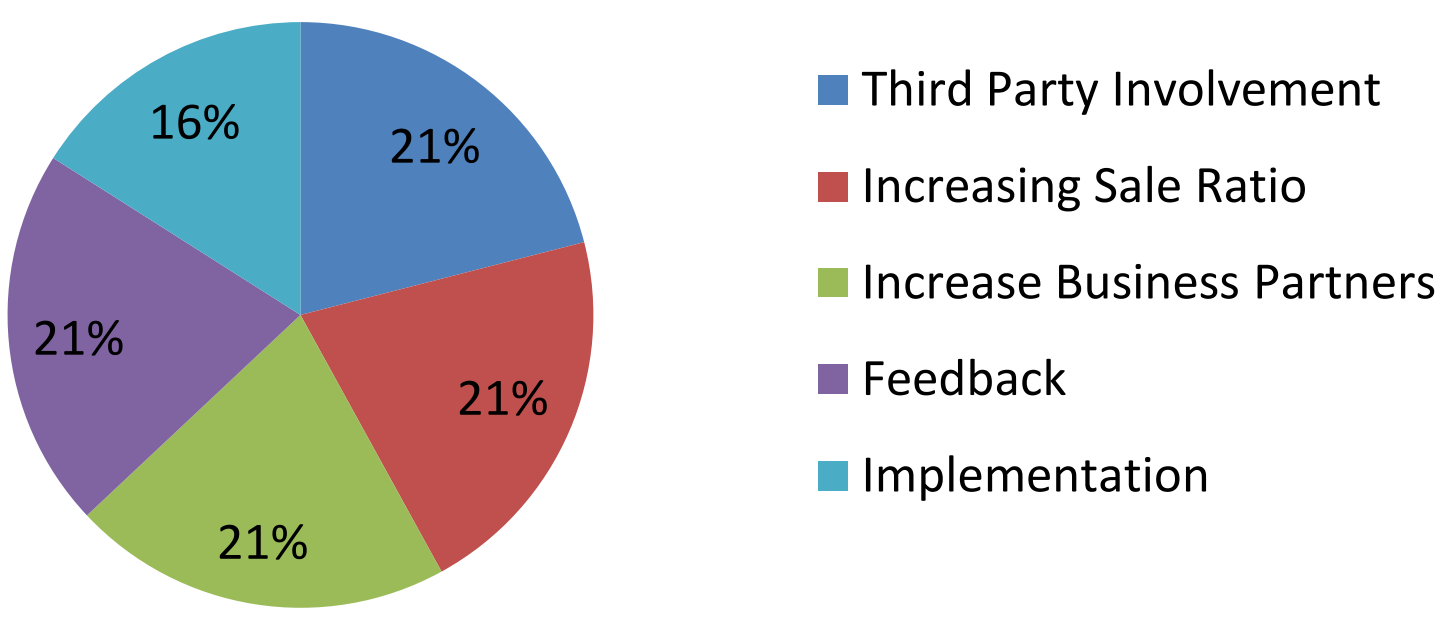

Figure 4 Perceived strengths of reputation based trust mechanism from the literature and industrial interviews

Reputation based trust mechanism provide valuable information to third parties [45][46][48]. According to the literature review and industrial interviews, reputation based trust mechanism increasing sales ratio and business partners [2][45][48]. Tracking feedback is another positive aspect of reputation based trust mechanism [45][47][48].

The identified weaknesses of reputation based trust mechanism have been classified into further factors on the basis of their similarities.

Table 5 Classification of the weaknesses in reputation based trust mechanism

| Factors | Weaknesses in reputation based trust mechanism |
|---|---|
| Honest Reputation | • Many reputation systems are too positive from seller's point of view<br>• A single entity can rate an item multiple times from many different identities<br>• Many people don't send any feedback and even don't rate at all after transaction<br>• Unfair rating are there to increase or decrease the product value |
| Cost | • Increase reputation by providing high quality low cost items<br>• Provide low quality services after getting high reputation<br>• Mostly positive feedback increased price but negative feedback don't affect them at all |
| Security | • After getting bad reputation, often sellers join the market with a different identity<br>• A single agent rate an item multiple times from many different identities |
| Bad Image | • The effects of Initial negative reputation cannot be removed even after getting many positive rating/feedback<br>• No information related to seller's business history is available |
| New Entity | • A new seller having no reputation at all is hard to trust<br>• Potential sellers face problems in the start |

Graphical representation of weaknesses associated with reputation based trust mechanism is given below:

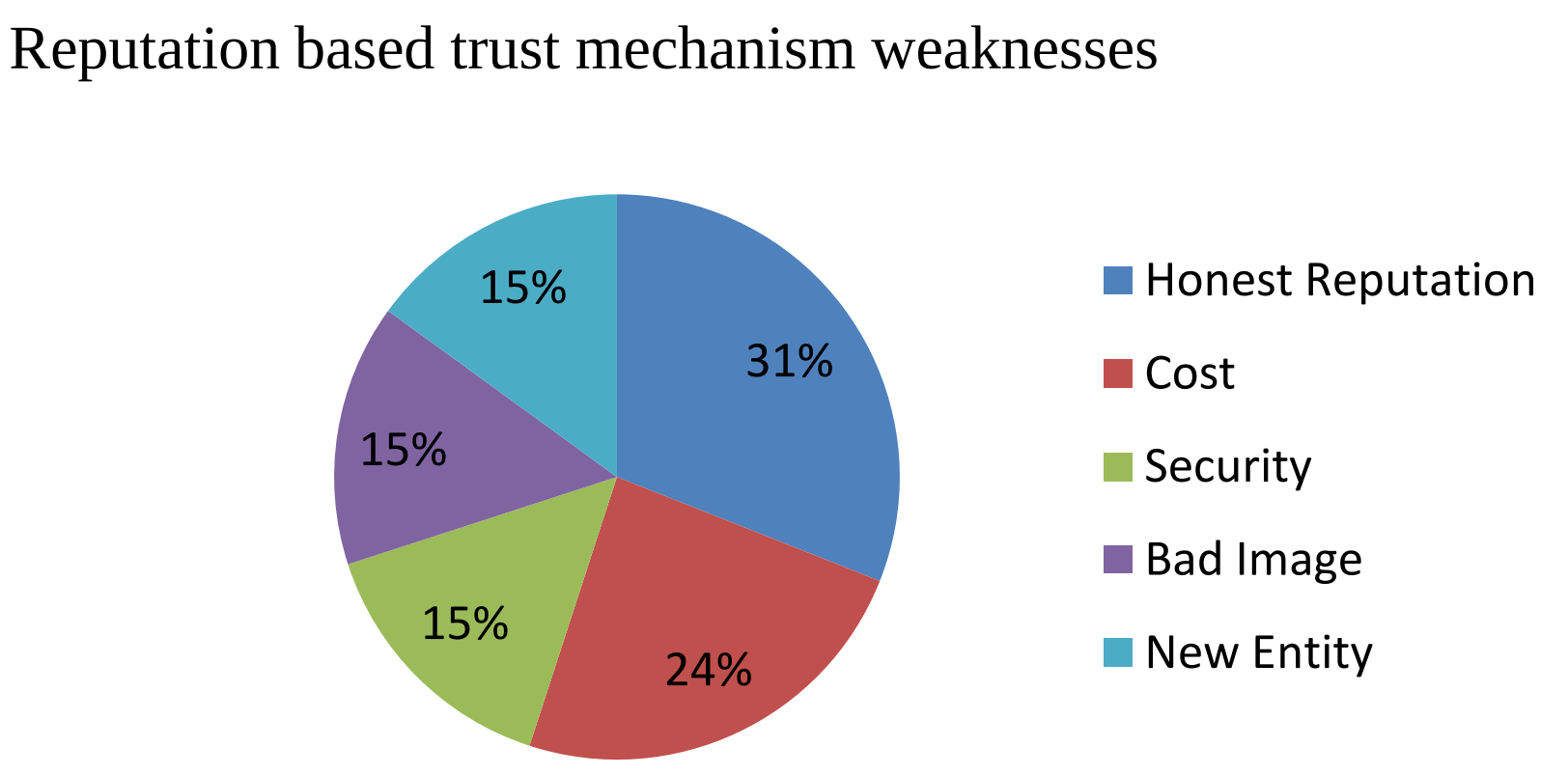

Figure 5 Perceived weaknesses of reputation based trust mechanism from the literature and Interviews

In the literature review, many reputation systems are too positive from seller's point of view and often buyers provide unfair rating. Cost of products and security threats are negative aspects of reputation based trust mechanisms according to the literature review and industrial interviews. The initial negative reputation cannot be removable and it's hard to trust on new seller with low or negative ratings [2][45][47].

## 3.3.1 Summery and Discussion

On the basis of findings from systematic literature review and industrial interviews, it is concluded that policy based trust mechanism are mostly used in the organizations where strict security requires. Strong binary decision (allow/deny) made under policy based trust mechanism, which allows only trustworthy clients to access specific resources details are in Section 3.1.6.1. Real world polices are often hard to implement, an interviewee suggests that it will be better to decompose these policies and then implement them. It will overcome the implementation problem along with these kinds of policies that can support new sellers in industry. Details are in interview 1, appendix B. On the other hand, reputation based trust mechanism are widely used in online e-commerce markets details are in Section 3.1.6.2. According to one of our interviewee, current online e-commerce industry does not have such a simple mechanism or policy that can support to new sellers.

The main purpose of conducting systematic literature review was to highlight the strengths/weaknesses of both policy and reputation based trust mechanisms. The industrial interviews were conducted to avoid researcher's biasness and to validate findings from systematic literature review. After conducting industrial interviews, some new aspects were extracted related to strengths /weaknesses of both policy and reputation based trust mechanisms which were not observed before in literature review. Another purpose to conduct industrial interviews was to get any possible solution/suggestions of problems from experts, which will be considered in the process of integration.

The identified strengths/weaknesses of both policy and reputation based trust mechanisms have been further classified into factors based on their similarities i.e. security, feedback, cost, bad image etc, described in section 3.3. On the basis of results from systematic literature review and industrial interviews described in section 3.3, a mapping process has been proposed in section 4.3.1. The identified weaknesses of one trust mechanism are addressed with strengths of other trust mechanism along with suggestions from experts.

Mapping process gives clear idea about the factors found from systematic literature review and industrial interviews, which involved establishing trust between two entities. The integrated trust mechanism has been proposed in section 4.4.2 that addresses main issues and their solutions of both policy and reputation based trust mechanisms. Furthermore, a part of the proposed integrated trust mechanism is implemented as a prototype and validated through an experiment in chapter 5.

# 4 PROCESS OF INTEGRATED TRUST MECHANISM

In this chapter, we present an integrated trust mechanism considering the strengths and weaknesses of both policy/reputation based trust mechanisms described in section 3.3. In first step we describe process being used in industry for both policy and reputation based trust mechanisms later we described our proposed integrated mechanism.

## 4.1 Trust Process of Policy Based Trust Mechanism

Policy based trust mechanism represents set of rules and strict security requirements used to verify the trustfulness of an unknown entity. Binary decisions made under policy based trust mechanism, if a user wants to use a service or want to get access he/she must pass though these kind of policies. Usually new user must have to provide set of credentials and set of policies determines user is trusted or distrusted. Decisions has been made under set of policies, these policies may vary on how much strict that particular environment is [10].

Declarative policies used under policy based trust to specify the access control decisions i.e. the requested services or resources should be allowed or denied [64].

### 4.1.1 Access Control in Policy Based Trust Mechanism

Trust negotiation process started when access resources owned by someone else. The outcomes of trust negotiations are to find sequence of credentials. All the credentials are dependent on policy. Credentials can only be disclosed when its access control policy has been satisfied. If the access is denied then no such credential sequence exists.

For example, an E-learning system is free for police in California State and a police officer wants to join that course. The steps involved in access control decision are described below [39].

**Step1.** Officer requests to access E-Learning system's free course.

**Step2.** E-Learning system required a driving license and police badge issued by California State Police which can prove requester is a police officer, and living in State of California.

**Step3.** Officer is agreed to disclose requested information, police badge is holding important information so officer wants to verify the E-leaning system belongs to Better Business Bureau.

**Step4.** E-learning system disclosed its Better Business Bureau membership card to officer.

**Step5.** Officer's trust level on e-learning system is increased by seeing Better Business Bureau membership card, now officer discloses police badge to E-learning system.

**Step6.** Officer is now validated by exchanging credentials with E-learning system. E-learning system allows police officer for free course.

The activity example of policy based trust mechanism is given in figure below.

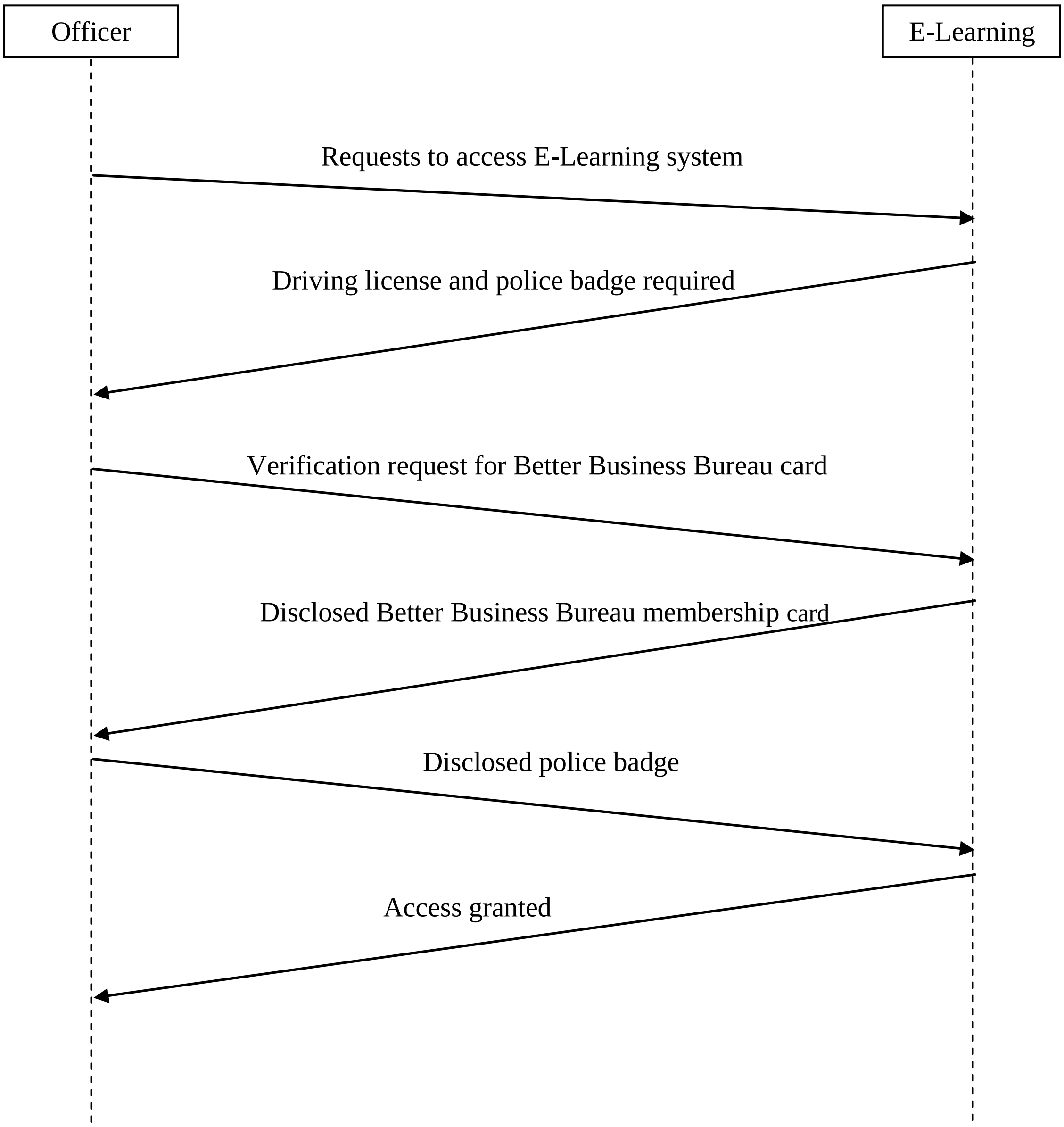

Figure 6 Activity example of Policy based trust mechanism

Each resource in this example can be an item associated with some credentials, some kind of repository with some set of rules that can allow or disallow the access to the system [39]. Every system has its own set of rules which can allow or deny getting access or getting a service from that particular system. The verification for an entity using authentication, authorization and certificate theory is helpful to increase trust but not sufficient [50].

### 4.1.2 Provisional Policies in Policy Based Trust Mechanism

Provisional policies address to certain decisions and requests or system may search itself on the behalf of certain credentials. Policy language should be able to handle the event condition action (ECA) rules approach [56]. In a response of a request a peer should be able to execute some action when ever this response can play its role to make negotiation successful i.e. Encoded business rules can be used for provisional predicates, inner rule allow discounts in particular session on low selling articles. An example of provisional policies is given below.

```
allow (Srv) ← . . . , session(ID),
in(X, sql:query('select * from low selling')),
enabled (discount(X), ID).
```

Figure 7 Example of provisional policies [56]

There are two set of rules if enabled (discount(X), ID) is false but the rest of condition is true then negotiator may want to enable second set of rule after that the overall rule becomes applicable. In case of second rule already enabled then no further action trigger. Dependent actions can be defined though metapolicy language, this language can only be applicable when an action needs to be triggered [56].

### 4.1.3 Third Party Certificate in Policy Based Trust Mechanism

The third party certificate theory is necessary but not fortified the need of satisfaction level on semantic web, because other aspects are necessary to consider in verification of trust in semantic web [50]. Policy based trust mechanism make decisions on some kind of properties. These properties may certify by cryptographic techniques, or with some kind of middle degree evidence and validation [56]. An integrated policy should mix these two kinds of policy form.

## 4.2 Trust Process of Reputation Based Trust Mechanism

Reputation based trust mechanism deals with personal experience and experience of others that supports to decide an entity trustworthy [65]. Past experience of an entity describes trustworthiness about another entity. That allows an entity to have trust, based on experience of others instead of the centralized trust management system.

Reputation based trust mechanism designed for distributed environment e.g. eBay (auction systems) [10].

## 4.2.1 Trust Level Calculation in Reputation Based Trust Mechanism

The degree of trust usually represents the trust level of an entity as a collective measure. The level of trust can be calculated in Boolean or in numeric values [51]. Weighted directed trust can be any real number between 0 and 1 [52]. The reputation of a particular entity calculated as, aggregation of all the numeric values (summation of individual rating) divided by number of ratters as a result in one percentage measure [52]. The formula is given in figure 8.0

$$\text{Reputation} = \sum_{i=1}^{n} wi \times \text{rating } i$$

Figure 8 calculation of reputation in auction systems [52]

Where $w_i$ represents weighted value of reputation ε [0...1] of a rater
$rating_i$ represents rating provided by a $rater_i$
n represents total number of raters who rated an entity

## 4.2.2 eBay Reputation Rating Calculation

There are many auction systems in industry every auction system has reputation management system of some sort. The eBay is best known auction system in industry. Our example auction system allows users to rate sellers and buyers at the end of their transactions. There are main three types of rating points that a user can use for other user [53].

- +1 point, feedback is positive
- 0 point, feedback is neutral
- -1 point, feedback is negative

Positive feedback represents transactions on auction system are smooth without any problem and level of trust (rating) is going high.

Natural feedback represents transactions on auction system are smooth but look like few problems with customer or buyer (depends on transaction nature) [53].

A negative feedback represents transactions on auction system are went though poorly.

Reputation calculated as, summation of distinct positive feedback minus summation of negative feedback in percentage with all the feedbacks.

Furthermore, eBay have star indications that means a user has reached up to certain amount of points which supports a user to increase the trust level of user. Based on eBay reputation system any user can decide whom to deal with by exploring seller's average points. Positive points will lead a user towards more business and users feel more trust to do business, against a person with too negative feedbacks or with a person who have no points on that particular platform.

# 4.3 Proposed Integrated Trust Mechanism

Reputation and Policy based trust mechanisms are being used distinctly for different environments. There is a need to integrate both trust mechanisms, which is able to fortify both trust mechanisms in order to overcome some weaknesses and get benefit of their strengths [62]. We propose an integrated trust mechanism having capabilities of both policy and reputation based trust mechanisms.

We have discussed and identified strengths and weaknesses of both policy and reputation based trust mechanisms in Section 3.3. These strengths and weaknesses were identified from literature review and from professionals working in industry. Detailed analysis of collected data can address different factors which are involved to enhance the trust.

In first phase of research, we used mapping process on factors perceived from strengths and weaknesses of both trust mechanisms. Mapping processes is a design mechanism which can capture real world problems and lead towards design a solution of that particular problem [54]. Tabular representation of mapping process used to organize and gather factors in order to improve their weaknesses, details are in section below. In second phase of research, we conducted an experiment to implement addressed factors in mapping process. Furthermore, results of the experiment will evaluate proposed integrated mechanism.

## 4.3.1 Mapping Process

In section below, mapping of weaknesses associated with reputation based trust mechanism to the strengths of policy based trust mechanism described in section 3.3 are highlighted. Identified weaknesses of reputation based trust mechanism can be addressed with strengths of policy based trust mechanisms along with solutions suggested from the expert opinions.

Table 6 Mapping of reputation weaknesses to the strengths of Policy based trust mechanism

| Weaknesses of reputation based trust mechanism | Solutions based on expert opinions and strength factors extracted from policy based trust mechanism |
|---|---|
| Problem 1.0<br>• Some reputation based systems are too positive (from seller's point of view)<br>• Single entity can rate multiple times<br>• Unfair rating which can effect product value | Solution 1.0<br>• Cross-rating, two entities rate each other<br>• Entities are not allowed to register with a different identity<br>• Only verified entities will be allowed to access |
| Problem 2.0<br>• Sellers can increase reputation with low cost products<br>• On high reputation, negative reputation effects a little<br>• The negative reputation can never fall, once given by other entity (or difficult process) | Solution 2.0<br>• Consider ratings between two entities as a unique, two entities can rate each other only latest rating will be used.<br>• cost of product should play its role in rating<br>• System saves most recent rating |
| Problem 3.0<br>• Re-entry in system with fake id<br>• Single entity can have multiple accounts<br>• Sing entity can rate multiple times from different locations | Solutions 3.0<br>• Access of resources are allowed after verification of credentials<br>• The approach for verification of credentials restrict duplicate accounts with all trustworthily relevant information (id, credit card etc) |
| Problem 4.0<br>• A new entity in system awarded no reputation | Solution 4.0<br>• Verification of credentials are also subject to have initial trust |

In section below, mapping of weaknesses associated with policy based trust mechanism with expert opinions to the strength of reputation based trust mechanism are highlighted. Some of the weaknesses associated with policy based trust mechanism can't fulfill with reputation based trust mechanism so we will propose some key factors from expert opinions.

Table 7 Mapping of policy based trust mechanism weaknesses to the strengths of reputation based trust mechanism

| Weaknesses of policy based trust mechanism | Solutions based on expert opinions and strength factors extracted from reputation based trust mechanism |
|---|---|
| Problem 1.0<br>• Irrelevant information asked in pre-registration phase, which is not relevant to required service | Solution 1.0<br>• After initial verification apply role based access control policies |
| Problem 2.0<br>• Binary decision system, where entities are not allowed to participate in decisions<br>• Static reputation, based on verified credentials e.g. 0 or 1 | Solution 2.0<br>• Reputation systems are flexible where reputation can change<br>• High reputation encouraged sellers to provide high quality services |
| Problem 3.0<br>• Implementation of policy based mechanisms are time consuming and directly related to object | Solutions 3.0<br>• Easy implementation even if it involves indirect repository<br>• Private communication is less between two parties |

#### 4.3.1.1 Discussion

The mapping process is based on factors defined in section 3.3 found from literature along with industrial interviews. These factors directly address to strengths/ weaknesses of reputation and policy based trust mechanisms. The strengths/ weaknesses acquired by conducting systematic literature review and industrial interviews. Further industrial interviews validated findings from systematic review and pointed some unknown strengths/weaknesses associated with policy and reputation based trust mechanisms.

Mapping process gives clear idea about the factors found from literature review and from industrial interviews, which are involved to establish trust between two entities. The mapping process and industrial interviews encouraged defining a process for integration of both reputation and policy based trust mechanisms.

# 4.4 Process Definition: Based on Mapping and Findings

## 4.4.1 Trust Classification

Based on previous study on trust, there are commonly two classes of trust i.e. direct trust and recommended trust [10]. Direct trust is an entity's own personal experience with someone else, about trustworthiness or usefulness in a particular domain. Recommended trust refers to the reputation information of another entity about trustworthiness or usefulness in particular domain [10].

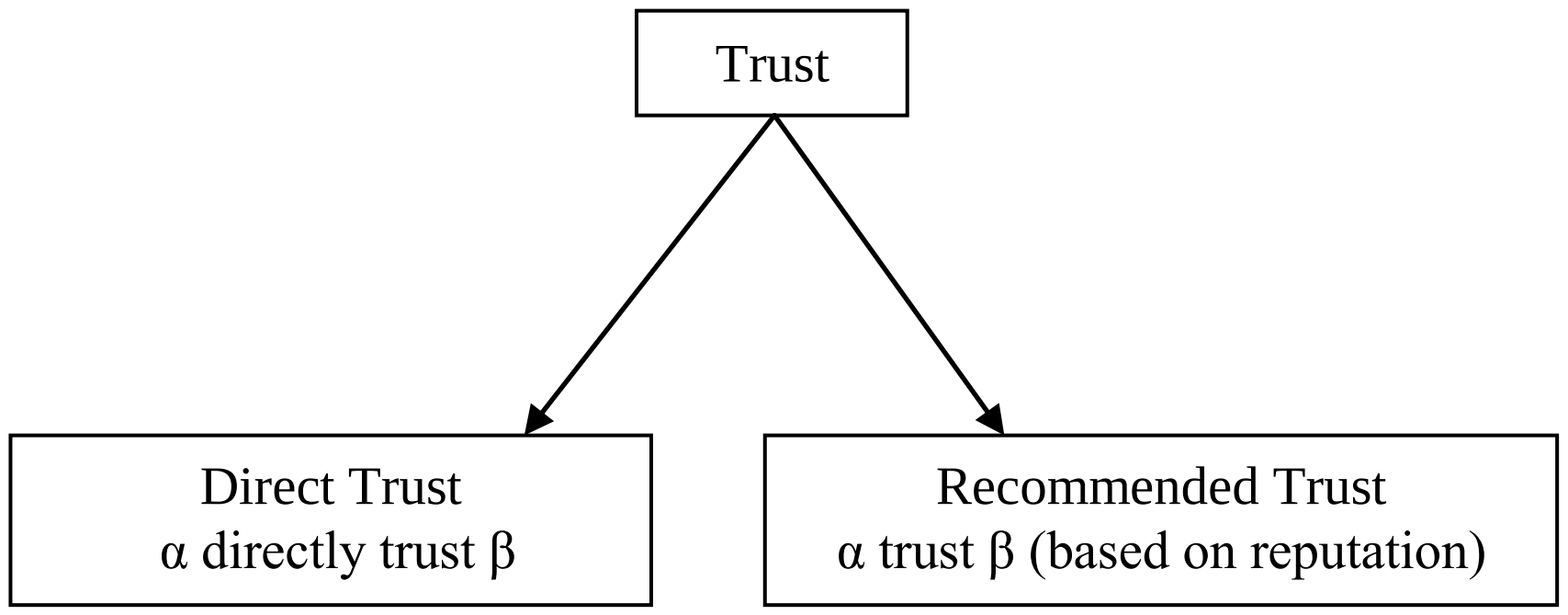

Figure 9 Direct and Recommended trust [57]

The integration of both reputation and policy based trust mechanisms perceived from previous work as mapping of weaknesses with strengths. Classification of trust defined above used as reference in our proposed integration along with the suggestions from industrial interviews as described in Section 3.3. The suggestions from industrial interviews were incorporated while refining integration process.

## 4.4.2 Proposed Integration Architecture

Software architecture refers to structural plan where development process used that plan as blue print describing different elements of system. Furthermore, it describes how different elements can be used to address the requirements of system. It helps at abstraction level to manage the development complexity of the system. The architecture's abstraction hides many complex details of the system [55].

### 4.4.2.1 Process Overview

Our proposed integrated trust mechanism is composed of three building blocks, the first block is based on the authentication of policy based trust mechanism and the third block is based on direct and recommended trust calculation which is directly related to reputation based trust mechanism. The real time calculation is involved in 2nd block, which is dependent on real time factors found from literature review and from the industrial interviews. The designed process is based on seller and buyer scenario in auction system. Verified buyer and seller can participate in proposed scenario. Trust opinion is a level of trust calculated from data repository and from real time calculations which will help buyer to decide which product of a particular seller is more trusted. If a seller is new in the market then policy based credential verification supports new seller to have initial trust level based on score or profile indication given by the repository.

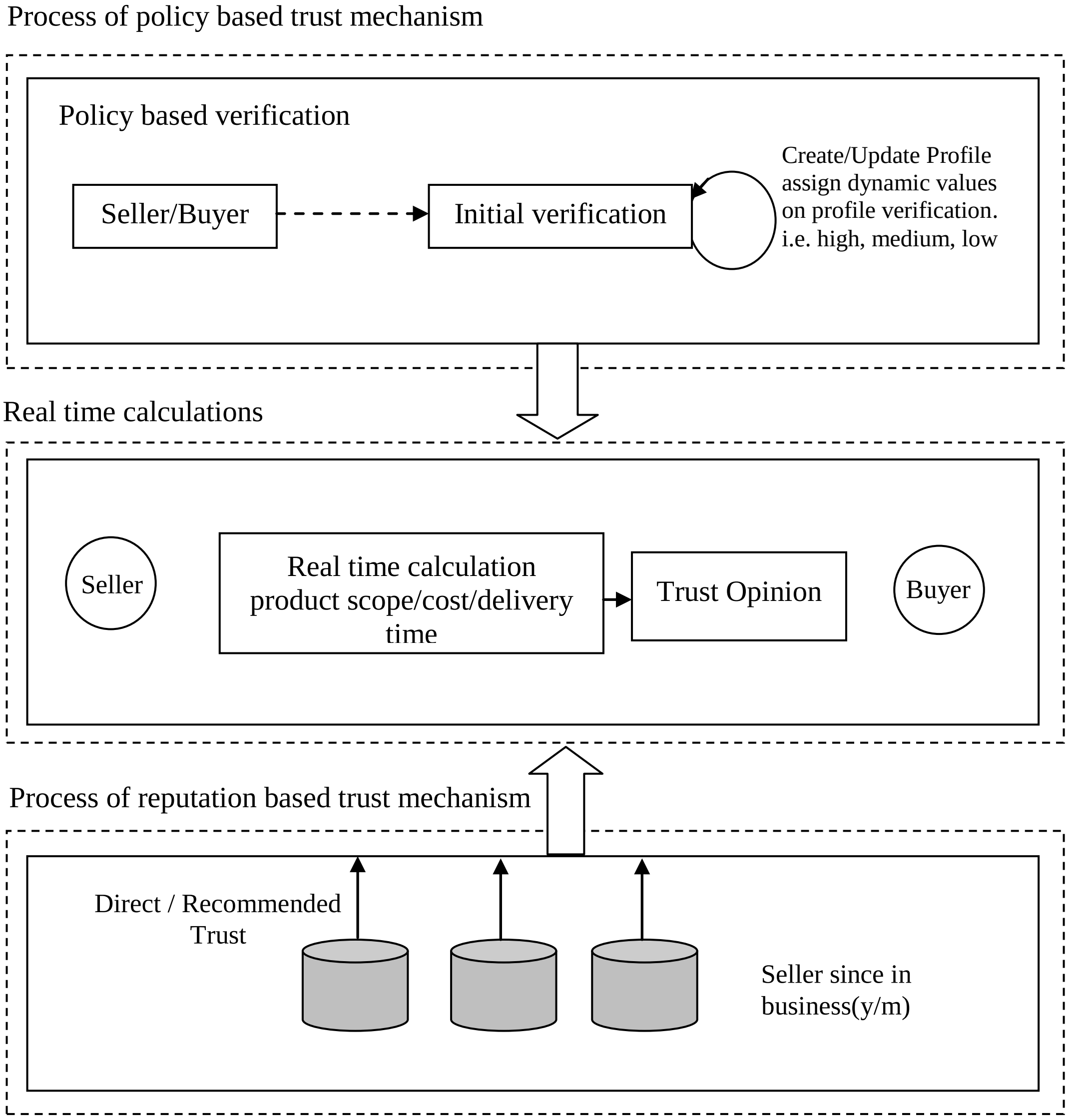

Figure 10 Proposed integrated trust mechanism

#### 4.4.2.1.1 Policy Based Verification Process

The first step of the proposed process is to verify the credentials given by an entity. One of the weakness factors was pre-registration phase asked irrelevant information which is not related to the required service, details are in section 3.1.6.1.2. We have tried to overcome this weakness in initial verification phase. Second major weakness described in mapping section as weaknesses of policy based trust mechanisms, was the binary decisions of policy based trust mechanisms, details are in section 3.1.6.1.2 and one interviewee pointed out this weakness and suggested to develop such a mechanism that could work like policy based verification and its output may similar to the reputation based trust mechanism. Suggested policy based mechanism for auction systems is described in figure 12

We can justify verification of credential process calculations semantically as

$$\text{seller(new)}\begin{cases} \text{verify(verified)} & \text{assign high} \\ \text{verify(some credential)} & \text{medium} \\ \text{verify (minimum)} & \text{assign low} \end{cases}$$

Figure 11 Proposed policy based decisions

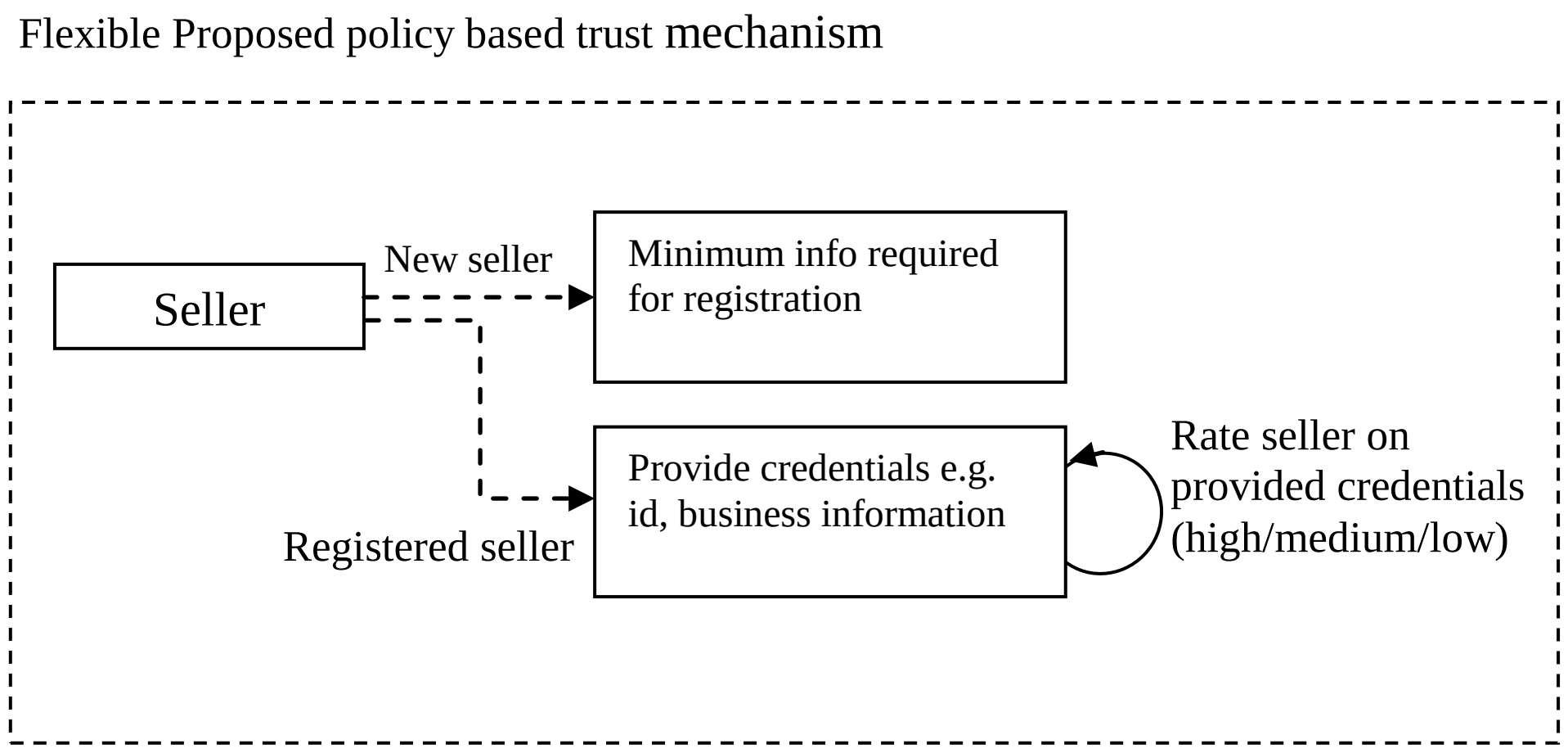

Figure 12 Flexible Policy based trust mechanism

#### 4.4.2.1.2 Reputation Based Process

The weaknesses of reputation based trust mechanisms fortified with the strengths of policy based trust mechanism. Identified weaknesses were fake, duplicate id and re-birth of a user addressed in section 3.1.6.2.2. Policy based credential verification is a strong shield among these kinds of weaknesses. The process of get rid from negative rating was not so simple or defined well in auction systems which is a big issue for potential seller, our proposed rating process have a capability to solve that issue as in cross rating and buyer/seller can rate each other once, data repository can hold only latest rating between two parties. Suggested detailed reputation based trust mechanism for auction systems is described in figure below.

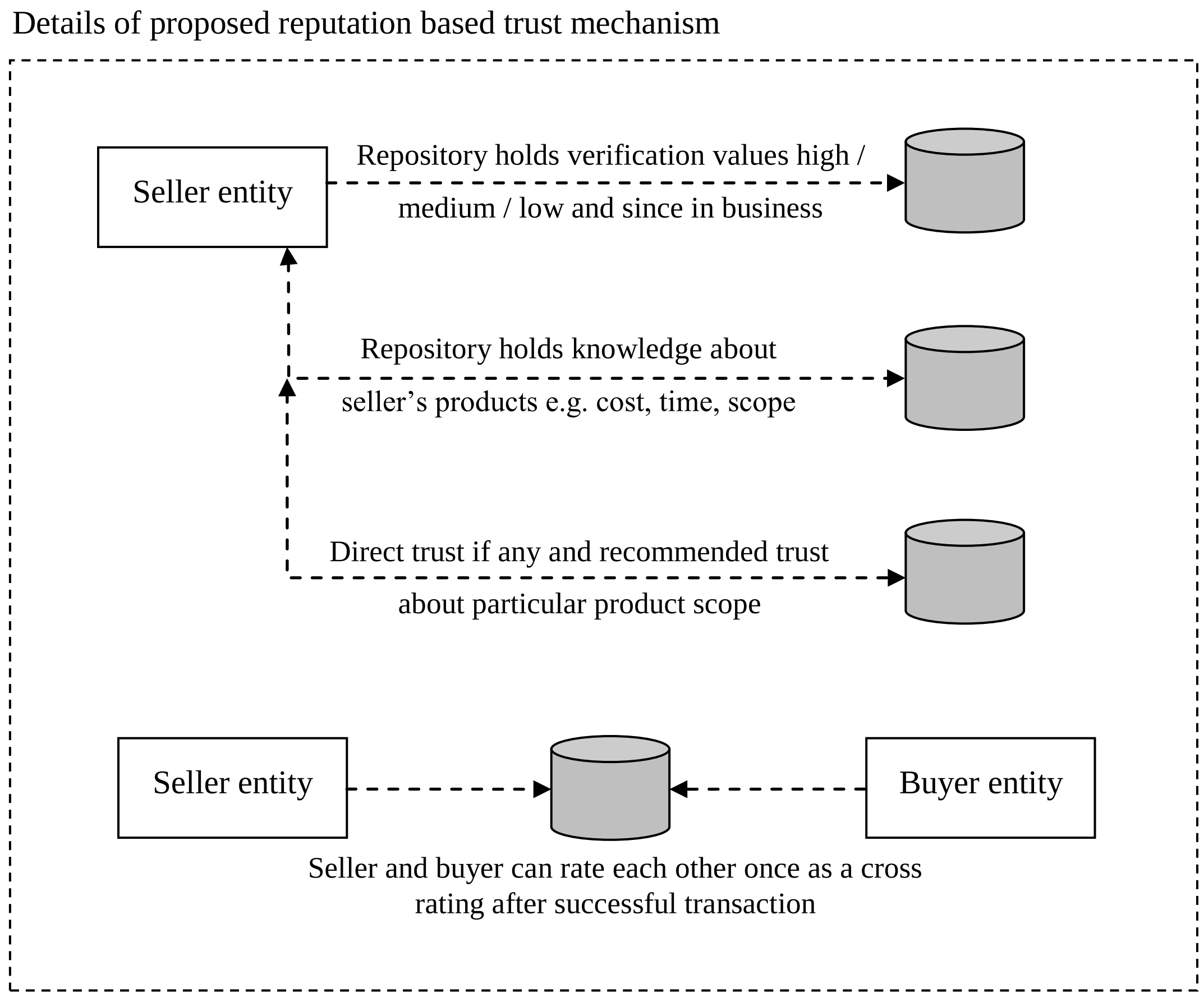

Figure 13 Detailed working of reputation based trust mechanism

### 4.4.3 Real Time Calculation Factors

There are two commonly used reputation calculation methods, approximate trust computation (ATC) where cache is involved to speed up the computation process. The second method is dynamic trust computation (DTC) where fresh data collection is made on runtime for further computation of trust values [66].

The proposed real time calculation is dependent on some factors identified in literature review and from the industrial interviews. In auction system scenario where buyers and sellers are directly involved doing transactions ranging between a few cents to thousands of dollars. There are some most important factors involved that can make an entity trustworthy or not [10]. In our proposed integrated trust mechanism the calculations are real-time based on these factors i.e. scope of product, cost of product, delivery time.

#### 4.4.3.1 Effect of Cost

The cost plays an important role in transactions. The cost of service or goods determines risk associated with that particular deal. For the micro payment of a product buyer may check only recommended reputation of seller but for high cost transaction the trust level would be different, so the cost is the major factor to make a seller trustworthy. Low cost of a product within a particular scope the trust level could be different and with high cost of a product within a different scope trust level could be different.

##### 4.4.3.1.1 Outcome

The outcome of involving cost factor in real time calculations is to quantify the trust level between low and high cost products. Involving cost factor gives clear idea about the seller's domain products and in which type of products belongs to that particular seller is trustworthy.

#### 4.4.3.2 Scope of Products

Dealing with transactions under auction system's platform, it is important to consider the scope of products in reputation repository system. If a buyer have good enough reputation in particular scope or domain of products then it's not necessary that he/she is good enough in second domain of products as well. Real-time calculation determined that the seller has enough ratings in that particular domain of products, thus system should take them into consideration. System will take scope of products as an input from reputation data repository.

##### 4.4.3.2.1 Outcome

The output, considering scope of product as an input from reputation data repository is to determine for which type of products a particular seller is trustworthy. If a seller doesn't have any reputation in repository about a particular domain of products then he will consider a new seller as an output of calculations involving factors in scope of products.

#### 4.4.3.3 Delivery Time

The availability and delivery time of a product is as important as cost of the product. If the delivery time is too long or the availability of particular product is not possible at required place then it’s useless for buyer and system will suggest buyer to avoid that product.

#### 4.4.3.3.1 Outcome

The delivery time factor will help buyer to avoid that kind of transactions which took you in more trouble after purchasing that product or service. That may belong to product delivery or belongs to services.

# 5 EMPIRICAL EVALUATION

In this chapter, we present an overview of empirical evaluation of integrated trust mechanism with experiment and analysis of the results. The experiments are mostly done in laboratory environment. The objective of conducting experiment in laboratory is to manipulate one or more than one variables. Furthermore, statistical analysis can be performed on the effect of manipulation [59]. The objective of empirical study was to determine the customer's trust satisfaction on sellers in auction system. Our main interest was to evaluate and compare newly developed real time trust opinions about new seller while browsing products of interest with eBay and Tradera. We need to determine where our designed integrated trust mechanism stands in relation of trust between customer and seller.

## 5.1 Experiment Planning

The experiment should be planned in some order and followed up, in order to control [59]. Sub section presents aspects related to experiment planning e.g. hypothesis, selection of variables, selection of subjects etc.

### 5.1.1 Hypothesis Formulization

It is important to know before conducting the experiment, what we intended to analyze in the experiment [59]. Our proposed integrated trust mechanism improves the trust of customer on seller particularly in case of new registered seller. The hypothesis may formulate as, the trust level of customer against new seller is same or increased between proposed trust mechanism, eBay and Tradera which will answer our RQ3.

#### 5.1.1.1 Hypothesis Statements

Null hypothesis, $H_{0\ \text{Support for new seller:}}$ The trust level of customer against new seller is same between proposed trust mechanism, eBay and Tradera.

Alternative hypothesis, $H_{1\ \text{Support for new seller:}}$ The proposed trust mechanism increased trust of customer on new seller against eBay and Tradera.

### 5.1.2 Selection of Variables

The variables are referred to as attributes/characteristics of an organization or individual that may be observed or measured within organization [15].

#### 5.1.2.1 Independent Variables

The independent variables are selected for this experiment is eBay, Tradera and proposed trust mechanism. In other words treatments are selected as independent variables.

#### 5.1.2.2 Dependent Variables

The decision of user is declared as dependent variable. Trust mechanisms are providing different trust opinions while selecting or browsing different products which lead a customer to make a deal in the form of transaction.

#### 5.1.2.3 Context Variables

The context variables are selected and described in section below, which are based on framework provided by Berander [58].

#### 5.1.2.4 Environment

The experiment was performed in controlled environment of personal machines. Where client/server environment was prepared and developed mechanism was installed on server for experiment.

#### 5.1.2.5 Subjects

In this experiment the subject were students, studying in master degree at Blekinge Institute of Technology. A total number of subjects were 20, who participated in this experiment. In order to ensure, all the participants have same level of understanding and each participant was briefed before conducting the experiment.

#### 5.1.2.6 Study Setup

The client/server based tool was developed to gather the data in repositories. The simple graphical interface was required for understanding of participants. The web based tool was selected to develop because changes are easy in these kinds of tools and these tools provide simple and improved interface. Every participant performed the experiment individually.

### 5.1.3 Selection of Subjects

The selections of subject were based on convenience sampling. The most conveniently and easily available participants were took part in this experiment [59]. Twenty subjects from Blekinge Institute of Technology took part in this experiment. The level of understanding with the auction system was necessary. The selected subjects were having same level of understanding with auction systems and minimum requirement was that they had experience with eBay and Tradera. Furthermore, to understand the experiment a training session was conducted with every participant before the start of experiment.

## 5.2 Experiment Design

The preparation is required during the design of experiment, conducting an experiment is not a simple task [15]. The design of experiment plays an important role in the success and to draw meaningful conclusions of experiment as an outcome of results. The results and outcome may affect if the design of experiment is not properly addressed.

The experiment has been conducted in a controlled environment to evaluate the assumptions as hypothesis. The implementation of proposed integrated trust mechanism as whole was not possible in available time and resources. Limited part of experiment was conducted as a web tool in seller/buyer scenario to address the RQ3 which is directly related to new seller's trust relation with customer. A set of tasks were performed by each participant and at the end set of questions were asked to gather numerical data, in order to verify that the proposed integrated trust mechanism is beneficial in buyer/seller scenario. The major issues found from literature review and industrial interviews were mapped (details are in section 4.3.1) and also considered in experiment i.e. support for new seller, bad image (negative rating), effects of cost, reputation calculation and registration policies.

In first step, every participant was asked to get register in system which is related to policy based trust mechanism. The implementation of registration phase was to reduce the problems in pre-registration phase, details are in section 3.3. The second step is about to analyze the different trust opinions involved real time calculation. Browsing different products of interest and find out different trust opinions (pervious trust score, profile score, recommended score), these scores personal/recommended are discussed in section 4.4.3. Furthermore, to enhance and improve the knowledge of participant about proposed trust mechanism we suggest them to browse same products without registration and asked them to analyze personal and recommended trust opinions. The next step is involved new seller's products along with new seller's trust opinions. The proposed system supports new seller and to remove previous bad experience in terms of bad rating. The last step was about to make a deal and give rate which works as cross rating and only latest rating will consider into account as discussed in section 4.4.2.1.2. The participant has bad experience with seller so a new deal with same seller can remove bad image of both buyer and seller. The calculation of real time reputation involves many factors, discussed in 4.4.3 which describes seller's recommend and direct trust with customer.

## 5.2.1 Experiment Design Type

The experiment is designed to investigate the trust level of customer on seller, particularly on new seller. In our experiment, comparison involves *"decision making"* of customer particularly in case of new seller. Proposed trust mechanism increase the trust level of customer on new seller to take a decision in making transaction or business.

The experiment design type may selected based on number of factors and treatments. We considered support for new seller in auction system as a factor along with comparison between eBay, Tradera and proposed trust mechanism as treatments. The suitable design type for our experiment is *"one factor with more than two treatments"* and the Likert Scale will be used for collection of data from experiment [60]. At the end of experiment, a set of questions asked to every participant along with every separate treatment and the answers were scaled with the help of likert Scale. The non-parametric test on likert scale data is most suitable for hypothesis testing [69]. There are two analysis technique suggested in book *"Experimentation in Software Engineering"* Kruskal Wallis and Chi-2 as a non-parametric test [59]. The nature of designed experiment one factor will be analyzed for hypothesis testing then the suitable data analysis technique is Kruskal Wallis test.

### 5.2.1.1 Instrumentation

The instrument of experiment i.e. guide lines about the experiment, forms, client machine, server machine with running some set of applications and developed prototype included in planning phase of experiment. A client machine was used to make real environment as auction system which access the server by sending/receiving HTTP requests. The integrated trust mechanism implemented in a prototype running on server machine, every participant dealing with client machine was actually working with integrated trust mechanism. At the end of experiment questionnaire forms provided to every participant to get the response of participant.

## 5.2.2 Validity Threats

During the data collection, validity threats may affect on validity of collected data. In order to validate the results a validity check must be done on collected data. The description about the threats and how they handled in experiment is listed below.

#### 5.2.2.1 Internal Validity

Internal validity threats are related to experiment execution that can affect the experiment if the researcher not aware of these threats.

*Knowledge about auction system*: The knowledge about auction systems and especially about the policy and reputation based trust mechanisms is necessary. Because, if a subject is unaware of these two mechanisms than it would have affect the out-come of the results. To overcome this threat we are restricting participants who do not have knowledge about the auction systems and their reputation systems.

*Instrumentation:* In a poor designed experiment, irrelevant questions asked to subject may affect the results. To cop this threat a pilot testing of experiment was done by the author himself.

*Time constraints:* Time can effect on the results of experiment if they were asked to perform the tasks in limited time. To cop this threat participant were asked to take their time as much as required.

#### 5.2.2.2 External Validity

Threats related to external validity deals with factors that can affect to generalize the results of experiment outside the experiment environment. The external validity may affected by the design chosen for experiment.

The experiment is designed to work in controlled manner with the limited participants. Limited part of the proposed trust mechanism developed for experiment. The developed part can be test with industrial data e.g. working auction systems but it may effect on experiment results. The outcome of experiment may differ with current results.

#### 5.2.2.3 Construct Validity

The application used in experiment was a limited developed part of proposed trust mechanism, which was inspired by eBay’s research lab. That developed prototype used to gather the participants trust response about new seller. Developed prototype may used with other treatments the outcome may differ from selected treatments. If a subject knows what is our intention to figure out then this can lead towards wrong or interrupted conclusion of experiment. There is a threat to the calculation as an outcome of experiment this threat can be minimized by not disclosing the purpose of experiment to subjects of experiment.

#### 5.2.2.4 Conclusion Validity

The conclusion validity may refer to as statistical conclusion validity. The experiment is designed to compare the trust of customer increased or still same after using designed trust mechanism for new seller in auction system. There is a chance that erroneous data may collect at the end of the experiment. To cop this threat we performed Kruskal Wallis test to calculate the significant difference between the mean values of treatments.

## 5.3 Experiment Execution

The experiment execution discusses different steps which were involved in performing the experiment. The objective of conducting experiment is to evaluate our designed part of integrated trust mechanism. To what extent the proposed integrated trust mechanism is helpful to make right decisions for the customers. The screen dumps of implemented prototype are included in the Appendix D. As the subjects were already experienced with Tradera and eBay auction systems, we focused conducting experiments with our developed system. In the end, we asked them questions about proposed integrated trust mechanism, eBay and Tradera to compare the results.

### 5.3.1 Experiment Operation

The experiment dealt with the system where a customer searched out for a product and the proposed system calculate trust factors between customer and seller. The preparation task were performed before conducting the experiment as committing with the participants, a pilot testing of designed experiment was done to make sure client and server environment is working as required. The screen dumps of implemented prototype are included in the Appendix D. 10 to 15 minutes introduction was delivered to every participant and the introduction was just for their understanding about the experiment.

Each participant performed set of task described in section 5.2 individually. The registration phase was required for participants who were participating in experiment. The registration phase was necessary to determine, to what extent a policy based registration effect on trust relation between buyer and seller. A sub task was performed before to make a deal between seller and buyer to browse products of interest and analyze the trust factors without registration on a system. After that, participants were asked to browse same products of interest with registration and then analyze the trust opinions. Whenever buyers browse the system then they can view various products with trust opinions of their owners in relation to the buyers. Whenever buyer make a deal with a system, the buyer notified about the contact information then asked to rate. Another situation was created to remove the bad image of seller with buyer. It was assumed that a buyer had a bad experience with seller so how bad image be removed, only latest reputation values will consider to calculate the personal and recommended trust described in mapping process in section 4.3.1. The trust factors are presented both in numerical values and in text values with icons. The trust factors were evaluated in real time calculations, through a wrapper called PHP script that generate html pages.

In the end of experiment, there were a set of questions and each participant asked to answer the questions related to their tasks in experiment. The entire questions are presented in Appendix C. Likert scale was selected to design and scale the questions, table of scale is presented in Appendix C.

# 5.4 Distribution of Questionnaire

After conducting the experiment, questionnaire related to experiment was handed over to participant. The questionnaire was about to measure the satisfaction level of participants between our proposed trust mechanisms, eBay and Tradera which are the most popular auction systems in the market. We have categorized every task related to factors, these are the same factors which we found from literature review and interviews, details are in Section 3.3. The interpretation of collected data will be presented in tabular and graphical forms. Questions were formulated to ask participants about eBay and Tradera as compared to proposed integrated trust mechanism, details about questions are in Appendix C.

Questionnaire formulation with trust factors used in experiment

Table 8 Group of questions against each factor

| Evaluation Criteria | Questions asked |
|---|---|
| Enhanced Policy/Security | • I found to have multiple accounts on system is impossible<br>• The policy of system supports new seller |
| Reputation Calculation | • I found trust opinions/information while browsing products is convenient<br>• The trust values shown with registration are not different than without registration |
| Cost Effects | • System supports cost and product category dynamically to have more trust on seller<br>• The cost of product effects reputation, low cost increases low and high cost increases high rate in reputation calculation |
| New seller | • My general intention to make deal with new seller is high/changed<br>• To what extent you can make a deal with new seller |
| Bad Image | • The system have a simple mechanism to remove bad image as a bad rating |
| Customer satisfaction | • I would recommend same proposed system for other auction systems<br>• Performance of the application was satisfactory |

The analysis of collected data was performed on the basis of Likert scale. The scale of questions was based on Strongly Agree, Agree, Less Agree, Disagree and Strongly Disagree [60] presented in Appendix C. The graphical representation of the results eBay, Tradera and proposed trust mechanism are shown in the figure 14.

## 5.4.1 Graphical Representation of Collected Data

The data was collected after execution or operation of experiment. The analysis of results was made on the basis of the collected data. Likert Scale was used for interpretation and measure participant's attitude by asking questions, answers of the questions determine to what extent participants are agree or disagree with a particular treatment. The graphical interpretation and representation of collected data for each treatment (eBay, Tradera and proposed trust mechanism) is given below.

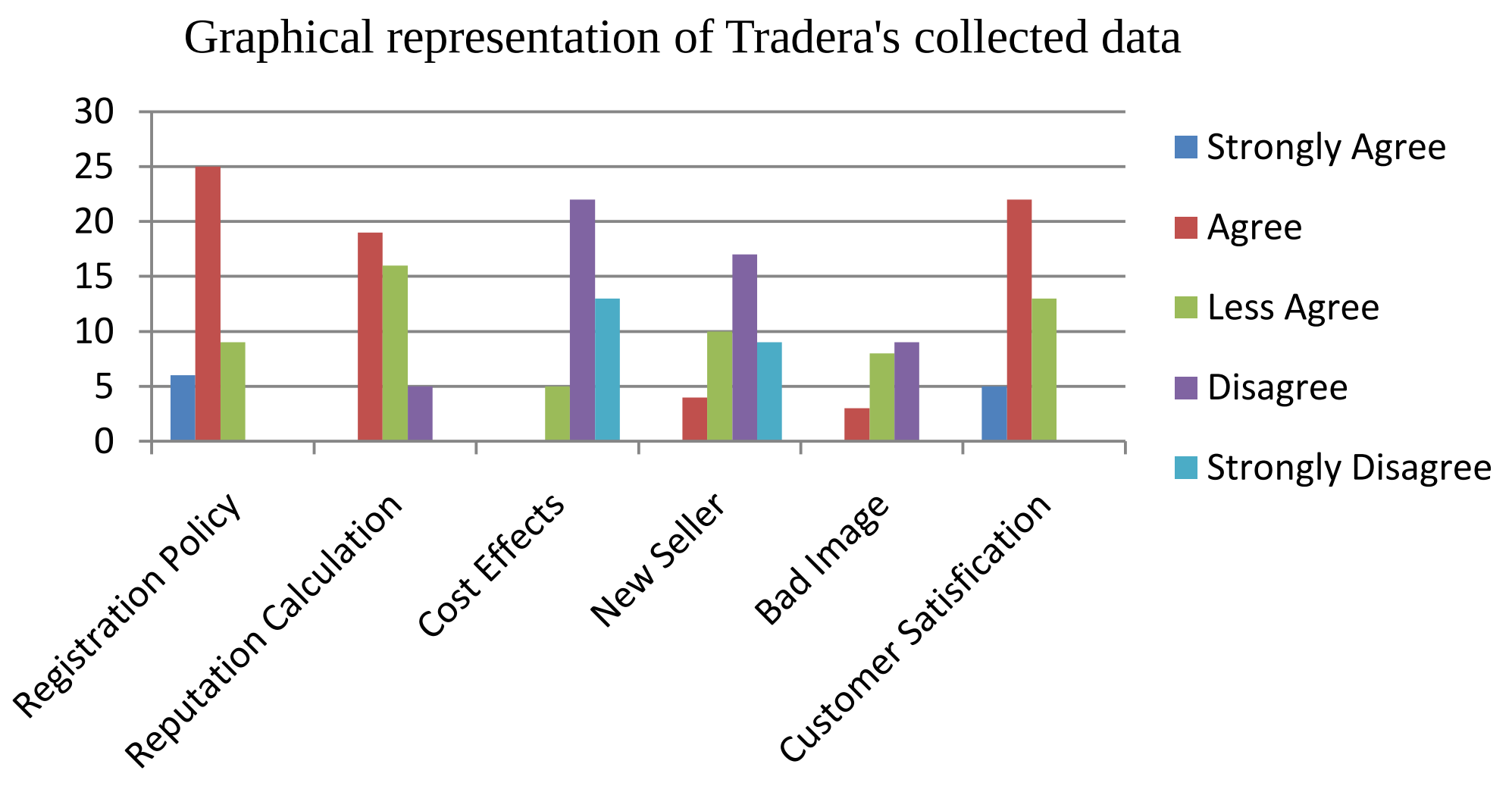

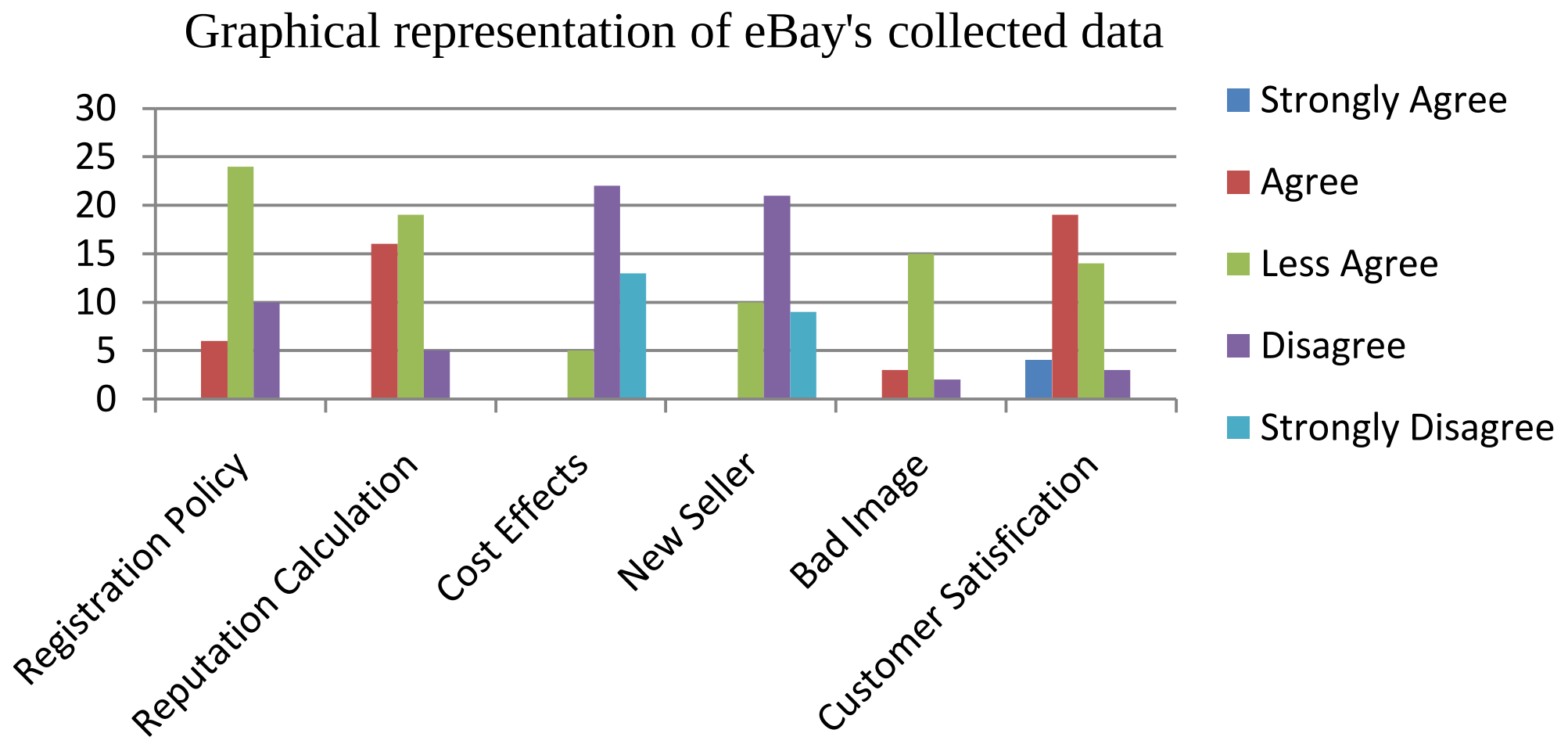

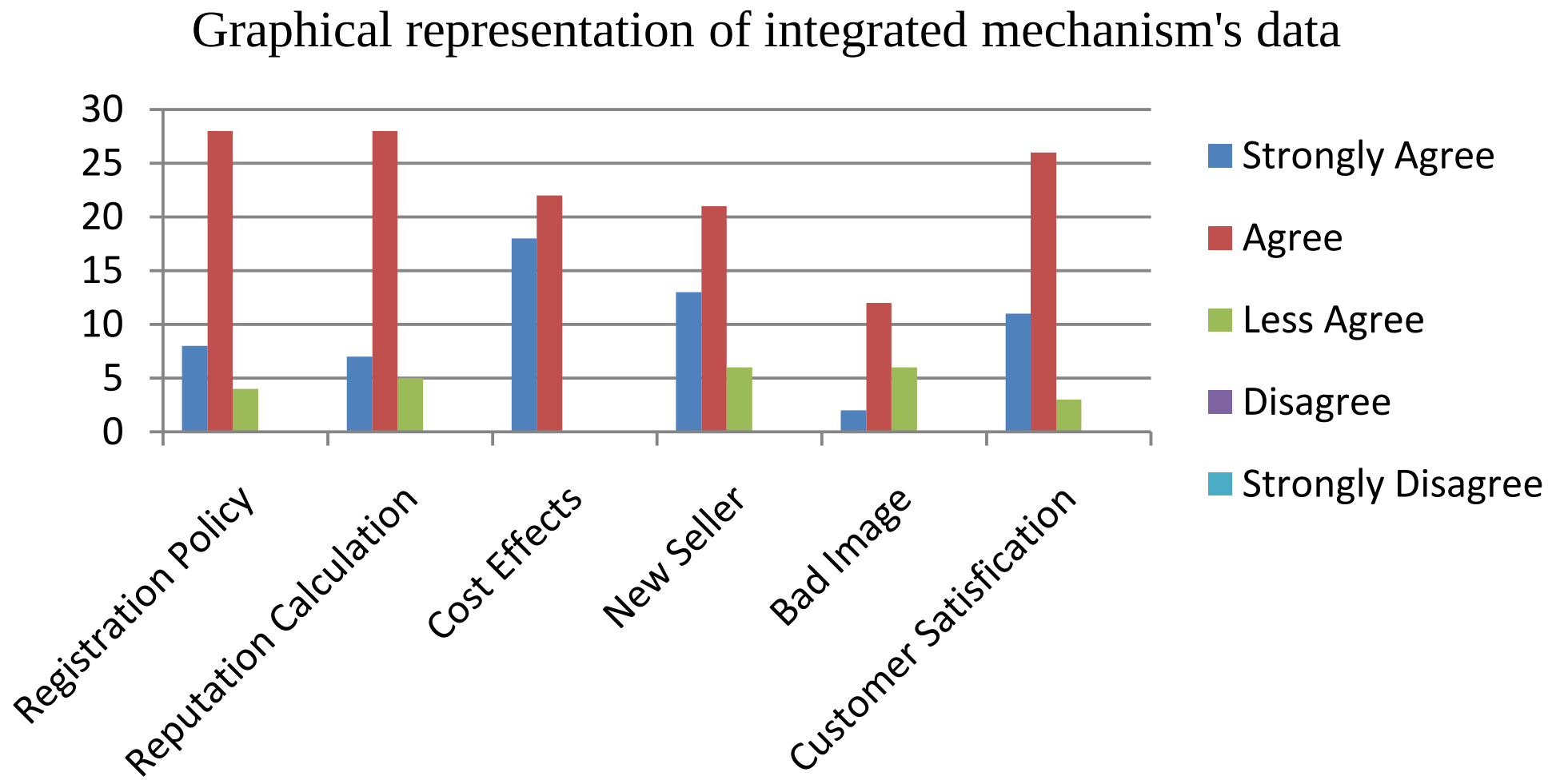

Figure 14 Graphical representation of the results eBay, Tradera and proposed trust mechanism

## 5.5 Results and Analysis

In this section, results of collected data are presented. The statistical analysis with interpretation of results is given on the basis of collected data from experiment. Interpretation of data is carried out by descriptive statistics along with hypothesis testing [59]. We used frequency and relative frequency measurements for descriptive statistics. The Kruskal Wallis a non-parametric test is suitable in the situation where ANOVA normality assumptions may not apply. As a hypothesis test a Kruskal Wallis was selected to draw some sort of conclusion as an outcome of experiment.

The main reason of hypothesis test is to address the RQ3 which directly related to support for new seller in auction system. The frequency tables which were made for descriptive statistics also used to calculate the mean [59].

The suitable test for comparing more than two treatments (eBay, Tradera, Proposed mechanism) using one factor (Support for new seller) is ANOVA, Kruskal Wallis and Chi-2. The output of experiment is based on likert scale data and any non-parametric test is suitable for this kind of data [69]. Kruskal Wallis test has been selected for hypothesis testing as a non-parametric test on likert scale data.

## 5.5.1 Statistical Analysis of Collected Data

The section below explains the difference between proposed trust mechanism as compared to eBay and Tradera. The different factors are discussed in section below which were used in experiment. A general view of descriptive statistics used before carrying the test, in order to understand the nature of collected data [59]. More details of frequency/relative frequency calculations are presented as frequency tables in Appendix C.

### 5.5.1.1 Registration Policy

There is much strength and weaknesses involved in registration policy, details are in section 3.3. An integrated trust mechanism has better registration policy in comparison with eBay and Tradera. According to the results of experiment shown in figure 14, 70% participants agreed and 20% strongly agreed on the registration process of proposed trust mechanism is more trust able. In contrast, 63% agreed and 15% strongly agreed in case of Tradera. The participants shown less trust on eBay's registration policies only 15% agreed while 60% less agreed.

### 5.5.1.2 Reputation Calculation

The proposed mechanism provide unique reputation calculation, every user can rate each other and only latest rating will be consider into account. Proposed trust mechanism involved with real time reputation calculation which involves product price, delivery time, and scope of product. We can explain this as if a person is trustworthy in computer related products then it's not necessary he/she is trustworthy in cloths dealing. The proposed trust mechanism supports these assumptions. The results shown, that our proposed trust mechanism is improved as 70% agreed and 17% strongly agreed. In contrast, 48% agreed on Tradera's reputation calculation and 40% participants less agreed. The agreeing ration is 40% and 48% less agreed with eBay's reputation calculations.

### 5.5.1.3 Effects of Cost in Reputation Calculation

The cost in an auction system at the time of deal is the most important factor which has been neglected by eBay and Tradera, cost factor is disused in section 4.3.1. Our proposed trust mechanism deals with cost value in reputation calculation furthermore, cost effects in rating at the end of a deal. High cost of deal can give high rating and low cost will give low rate in proposed trust mechanism. The results of experiment shown clearly 45% participants strongly agreed and 55% agreed that cost must effect on reputation calculation. In contrast, 55% disagree and 33% strongly disagree in case of Tradera and eBay.

#### 5.5.1.4 Bad Image (Negative Rating)

The target auction systems e.g. eBay and Tradera do not have a simple mechanism to remove bad image of seller or customer, in case of any one earn bad rating in a deal. The proposed trust mechanism has a simple and fair mechanism to remove the bad image of users. The proposed trust mechanism allows to rate each other and only latest rating will count in reputation calculation. Results shows 60% agreed, 10% strongly agreed on proposed trust mechanism for removal of bad image of any user. Other hand, 15% agreed, 40% less agreed in case of Tradera where as this percentage is 15% and 75% in case of eBay.

#### 5.5.1.5 Customer Trust Satisfaction

At the end, couple of questions asked to participants about the overall trust satisfaction of all three treatments working as auction systems. According to the figure14, 65% of participants agreed and 28% were strongly agreed on proposed trust mechanism providing overall trust satisfaction. On the other hand, 55% agreed and 13% strongly agreed on Tradera while 48% agreed, 10% strongly agreed on eBay's overall trust satisfaction.

#### 5.5.1.6 Support for New Seller

The proposed system supports new seller on a platform as compared to eBay and Tradera. Strong policy based registration determines that this seller is verified. According to the results, 33% strongly agreed, 53% agreed that the level of trust is increased on new seller using proposed integrated trust mechanism. On the other hand 10% agreed, 25% less agreed and 43% disagreed in case of Tradera. 25% less agreed and 53% disagreed in case of eBay.

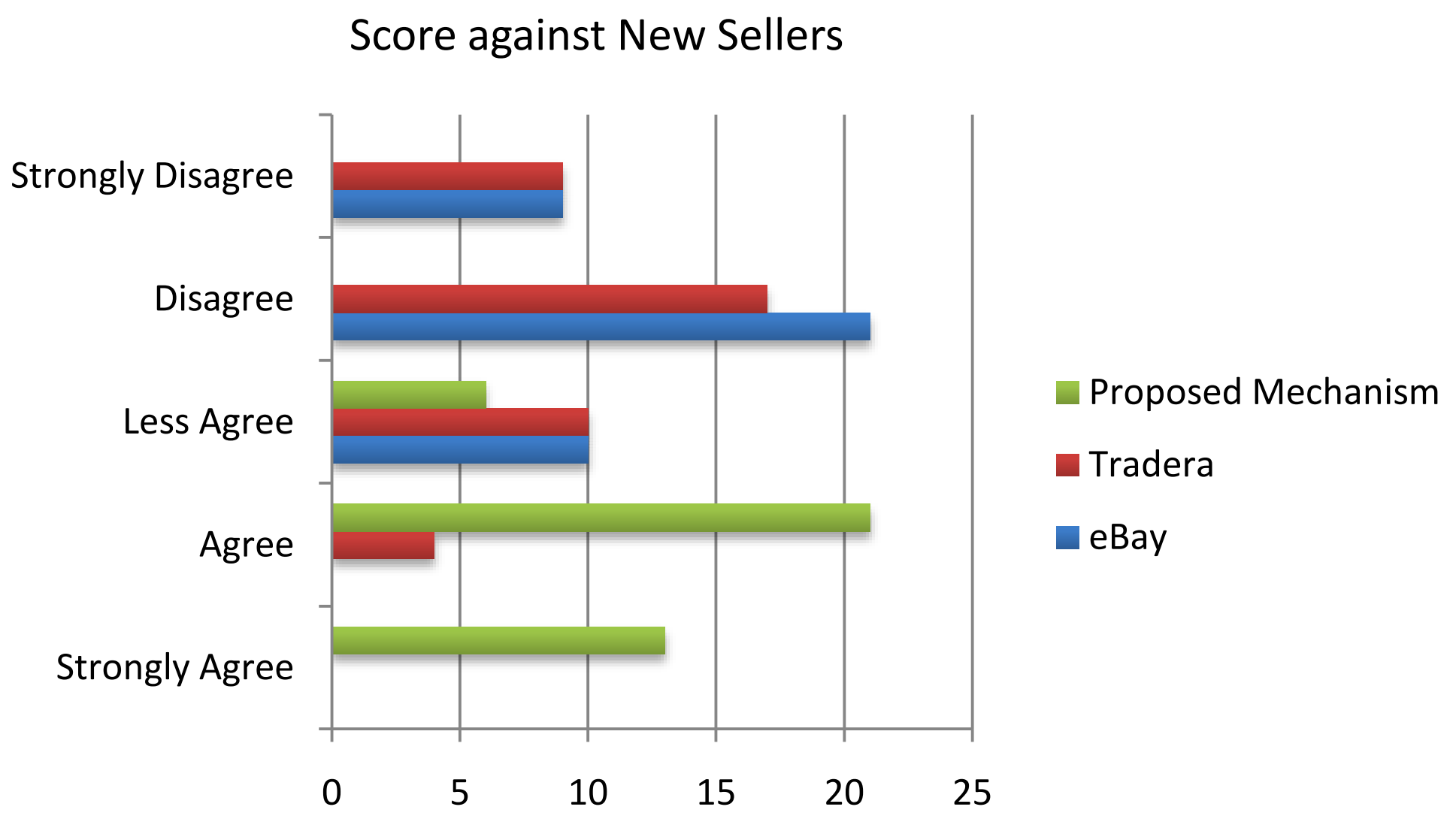

Figure 15 Comparison of new seller, details are in Appendix C

The *mean* calculation is presented for further analysis of data, collected as support for new seller from the experiment.

SUMMARY

| Groups | Count | Min | Max | Sum | Mean | Median | Variance |
|---|---|---|---|---|---|---|---|
| Proposed Mechanism | 40 | 1 | 5 | 167 | 4.175 | 4 | 0.455769231 |
| Tradera | 40 | 1 | 4 | 89 | 2.225 | 2 | 0.845512821 |
| eBay | 40 | 1 | 3 | 81 | 2.025 | 2 | 0.486538462 |

Figure 16 Statistical summary of factor support for new seller

Statistical analysis is composed of equal sample size of data i.e. “40” responses for each group. Greater no of sample size may have different results. There is a great difference between the *calculated mean* of proposed mechanism, Tradera and eBay. The value of median for proposed trust mechanism is “4” which described the average participants agreed and their level of trust is increased. The values defined against each scale is defined in Appendix C

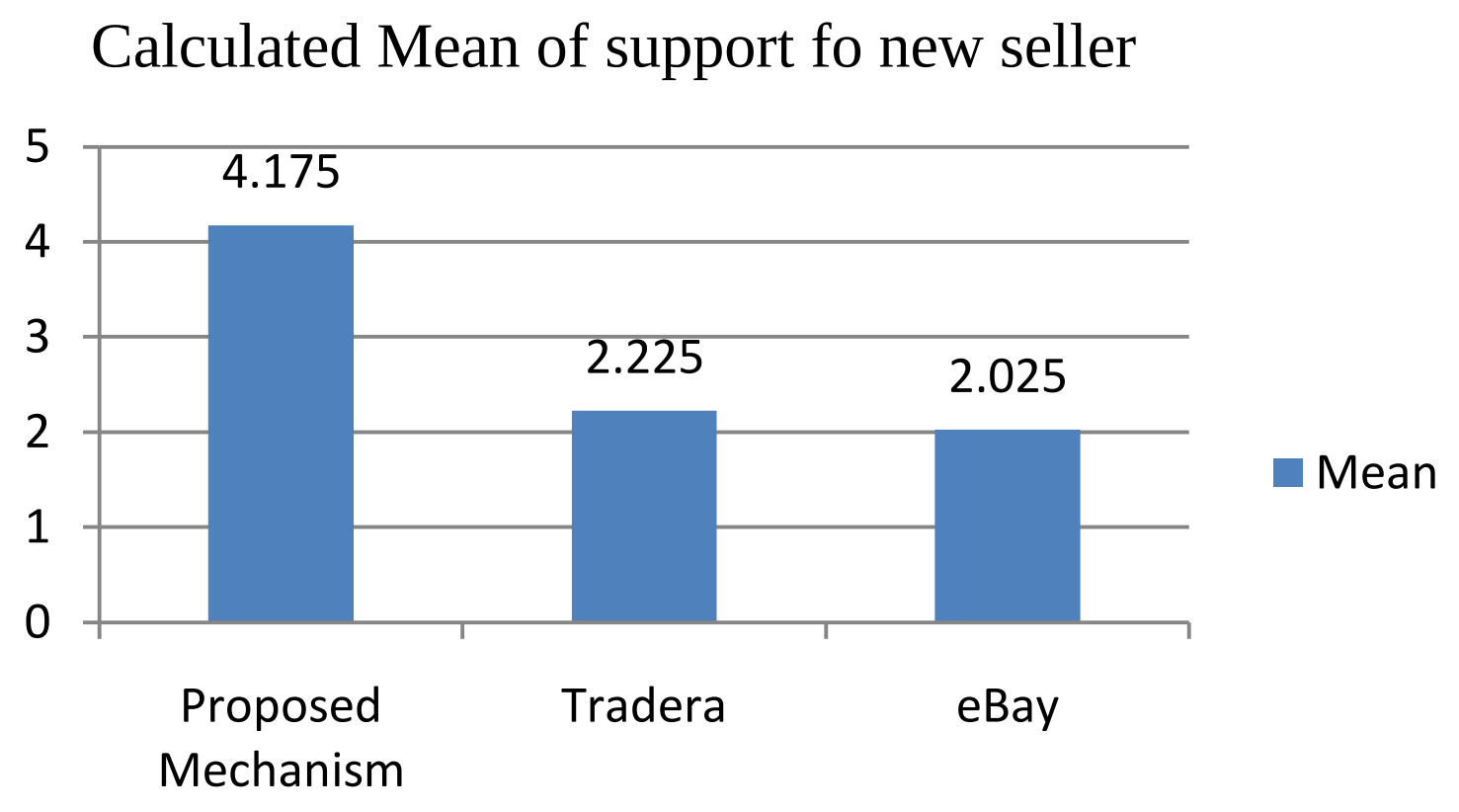

Figure 17 Mean comparison of three treatments

Comparison figure clearly displaying the proposed integrated trust mechanism is more supportive than eBay and Tradera. On the basis of statistical analysis we also conclude that the trust level of customer about new seller is improved against eBay and Tradera. The trust factor "support for new seller" is directly related to RQ3. We also used Kruskal Wallis testing for further analysis of hypothesis described in section 5.1.1.

### 5.5.2 Validation of Calculated Results

In this section, hypothesis test is performed for one measurement i.e. support for new seller. The main purpose of using statistics is to summarize collected data in clear and understandable form [59]. Kruskal Wallis test is performed in order to verify the hypothesis which addressed RQ3. The Kruskal Wallis test can be used if each sample size has minimum 5 in numbers [70]. The purpose of conducting Kruskal Wallis test is to cast interference between *calculated mean* of three groups. The hypothesis testing was done only with factor *"support for new seller"* and data table is given in Appendix C.

First of all we calculate the Degree of Freedom where "k" represents number of samples or treatments [70].

df = k -1

df = 3-1 = 2

The α typically set to 5% which is 0.05, the critical value = 5.99 found from chi square table with degree of freedom = 2 and α = 0.05. The chi square table-5 is given at the end of the book "100 statistical test" [70].

## 5.5.2.1 Calculate Mean of Ranks

In this section we calculate the sum of ranks by combining and arranging k samples and give them a rank number. The ties occurred in most of the cases with data collect from likert scales. In this situation we will calculate the mean value of available rank numbers. The calculated mean values against each rank number are given below.

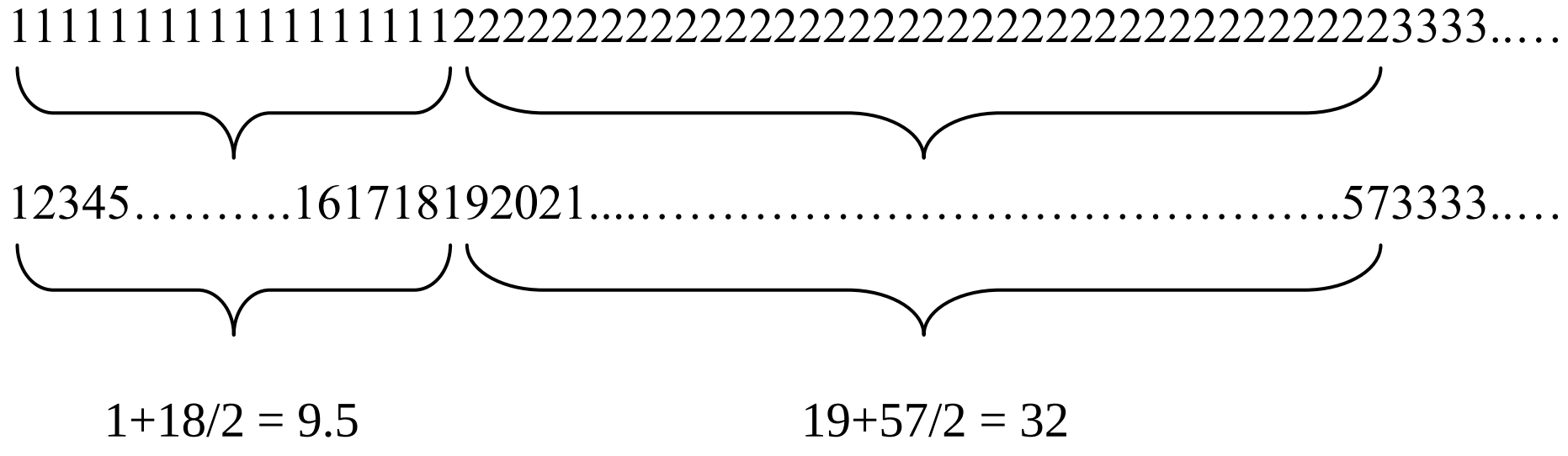

Figure 18 Mean value calculation example

The mean values are calculated by ordering all scale values then dividing first and last number by 2. The detailed calculated mean values are given in table below.

Table 9 Mean values of ranks

| Scale | Tradera | eBay | Proposed Mechanism | Mean Rank |
|---|---|---|---|---|
| No of 1s | 9 | 9 | 0 | 1+18/2 = 9.5 |
| No of 2s | 17 | 21 | 0 | 19+57/2 = 32 |
| No of 3s | 10 | 10 | 6 | 58+84/2 = 71 |
| No of 4s | 4 | 0 | 21 | 85+110/2 = 97.5 |
| No of 5s | 0 | 0 | 13 | 111+114/2 = 112.5 |
| N = 120 | n = 40 | n = 40 | n = 40 | |

According to the procedure described in [70] ties occurred in sample data then we calculated mean values of each number.

## 5.5.2.2 Sum of Ranks

The calculation of each rank value is presented below, which will use further in calculation of H value. There are three ranks in collected data with n = 40. The rank will calculated by multiplying mean value with number of occurrence or each scale.

*R1 represents Tradera

R1 = 9 × 9.5 + 17 × 32 + 10 × 71 + 4 × 97.5

R1 = 85.5 + 544 + 710 + 390

**R1 = 1729.5**

*R2 represents eBay

R2 = 9 × 9.5 + 21 × 32 + 10 × 71

R2 = 85.5 + 672 + 710

**R2 = 1467.5**

*R3 represents proposed integrated trust mechanism

R3 = 6 × 71 + 21 × 97.5 + 13 × 112.5

R3 = 426 + 2047.5 + 1462.5

**R3 = 3936**

#### 5.5.2.3 Hypothesis testing

The statistics test approximates chi-square distribution using degree of freedom which is k-1 = 2 each $n_i$ = {1, 2, 3} must be at least 5 for approximation to be valid. In our case n = 40 which is greater than 5, hence we can perform Kruskal Wallis test.

$$H = \left\{ \frac{12}{N(N+1)} \sum_{j=1}^{K} \frac{R_i^2}{n_j} \right\} - 3(N+1).$$

Figure 19 Formula for Kruskal Wallis test

$$H = \left\{ \frac{12}{(120)(120+1)} \; (1729.5)^2 + (1467.5)^2 + (3936)^2 \right\} - 3(120+1)$$

* N = 120 the total number of scale values.

H = (12/120(120+1) × (74779.25625 + 53838.90625 + 387302.4) - 3(120+1)

H = (12/14520 × 515920.5625) – 363

H = 63.38

The difference between three treatments is considered significant if the value of H is greater than the critical value which is 5.99.

We can interpret that, the trust level of customer regarding support of new seller's *calculated mean* is different against eBay, Tradera and proposed trust mechanism. Calculated H value is greater than critical value therefore, we reject the null hypothesis $H_0$ in favor of alternative hypothesis that is *"proposed trust mechanism increased trust of customer on new seller against eBay and Tradera"*. The descriptive analysis and hypothesis testing described, average participants are agreed that the trust level of customer has been improved on proposed integrated trust mechanism compared to eBay and Tradera.

# 5.6 Suggestions

There were many factors used in experiment to compare proposed trust mechanism with eBay and Tradera. Following points were observed while analyzing the results of experiment, described in figure 14. The trust factors analyzed in this thesis study might be helpful in improving the auction system's trust management.

- *Registration policies should be flexible, too strict and binary decisions may not be helpful in an environment such as auction systems.*

According to the results presented in section 5.5.1.1 people are much interested in easy and flexible registration policies. Proposed trust mechanism provides three different levels of profile/registration which can help sellers in a platform. Dividing registration policies in level supports new seller who initially do not have any reputation. The proposed integrated trust mechanism does not support sellers to declare them trustworthy but in terms to declare their level as verified seller i.e. high/medium/low

- *Reputation calculation mechanism should be changed; current mechanisms are using reputation as an average of user's transactions.*

The purpose of reputation calculation is to declare that the seller have enough experience in these categories of business. He/she have done enough satisfactory transactions with no complaints. The proposed trust mechanism is providing dynamic real time reputation calculation mechanism where real time calculations with factors are suggested. These factors are presented in section 3.3 and results are clearly describing that their behavior is changed where these factors are involved. The auction systems without factors have less interest of people in contrast people show more interest in proposed integrated trust mechanism where some factors were involved in real time reputation calculation.

- *Current auction systems may change their mechanism to remove the bad rating in result of unsuccessful transaction.*

Although, almost every auction system have a mechanism to remove the bad rating but all these mechanisms are too complicated or time consuming. They must adopt simple and quick mechanism to remove the bad rating. We can understand this scenario as, a seller was not much interested in transactions at the beginning but after a while he realized that he did some mistakes and awarded bad rating now he want to remove bad image and turned towards serious transactions. Unfortunately, current mechanisms have not shown serious intensions in this way. Proposed trust mechanism is successful in order to entertain the removal of bad image against eBay and Tradera. In addition, communication between seller and buyer to remove bad image might be required in real time scenario.

In the end, we conclude that many trust factors were found from theoretical studies which are described in section 3.3. Some of them were addressed in empirical study in available time and resources. There may be more work required or unseen logical errors may exist in proposed integrated trust mechanism. The work declared so far shown that the proposed trust mechanism has increased overall customer satisfaction in given scenario, details are in 5.5.1.5.

# 6 EPILOGUE

## 6.1 Conclusion

In this thesis, we introduced and validated an integrated trust mechanism to improve trust relationship between seller and customer in auction systems. In the first step, the strengths and weaknesses of policy and reputation based trust mechanisms were identified. This was accomplished by conducting systematic literature review with industrial interviews in order to investigate the strengths and weaknesses observed by professionals from industry.

The results obtained from first step showed that a gap exists among the perceived results of strengths and weaknesses in both trust mechanisms keeping theoretical and industrial perspective. It was also observed, there was no formal trust mechanism being used in industry that can support new seller in auction systems. The current trust mechanisms for calculating the reputation in industry is based on number of transactions, no matter what was the context of transaction and how much cost of product was involved in transaction. For these reasons an integrated trust mechanism was introduced to achieve benefit from both policy and reputation based trust mechanisms. In second step, the integration was defined which was based on mapping between identified strengths and weaknesses of both policy and reputation based trust mechanisms. Furthermore, the integration process was also based on suggestions of industry professionals, these suggestions were obtained from interviews.

In third step, the integrated mechanism was validated by conducting an experiment to compare identified factors against eBay and Tradera. However, minimum identified factors were selected for conducting experiment in available time and resources. This thesis study covers gap of both policy and reputation based trust mechanisms by introducing an integration of both trust mechanisms. The results show that the integration with involvement of factors increases the trust level of customer. The trust opinion supports customers to make decisions. Furthermore, there is no defined simple mechanism which support new seller and remove bad image of seller in eBay and Tradera. Result show that the participants of experiment have more trust on new seller in proposed trust mechanism against eBay and Tradera. In addition, the integrated trust mechanism also facilitates some more factors e.g. enhanced mechanism for reputation calculation, enhanced mechanism for registration.

Although both eBay and Tradera are working in industry but still there are chances for possible improvements where some of issues are addressed in proposed integrated trust mechanism.

### 6.1.1 Answers to the Research Questions

In the section below, answers are mapped with the relevant research question.

*RQ1: What kinds of circumstances are more suitable for policy respective reputation based trust mechanisms in auction systems?*

The RQ1 was answered by dividing this question in sub phases. In the first phase, strengths and weaknesses of both trust mechanisms were identified which were obtained from systematic literature review and industrial interviews details are in section 3.3. Many weaknesses and strengths have been identified. However, precise and most important ones are mentioned.

In second phase, the analysis of identified strengths and weaknesses has been presented in section 3.3.1 which addressed the most suitable circumstances of both policy and reputation based trust mechanisms.

*RQ2: How to integrate both reputation and policy based mechanism to increase chances of trust?*

The RQ2 was answered, based on conducting analysis of strengths and weaknesses presented in section 3.3. A mapping process was done using strength of one mechanism with weakness of other, details are in section 4.3.1. The mapping process gives us idea about the integration of both trust mechanisms, proposed integrated trust mechanism presented in section 4.4.2.

The validation of proposed integrated trust mechanism was done by conducting experiment along with formulation of hypothesis details are in 5.1.1. A part of proposed integrated trust mechanism was deployed to collect the data which satisfy hypothesis. In section 5.4.1 describes the graphical representation of collected data from experiment.

*RQ3: Could there be benefits of using both reputation and policy based trust mechanisms in establishment of new seller relation with customers in auction systems?*

The results of hypothesis testing along with statistical analysis illustrate that the null hypotheses in particular case of new seller in auction system is rejected. The detailed answer of RQ3 is presented in section 5.5 along with hypothesis testing in section 5.5.2 some suggestions presented in 5.6.

## 6.2 Future Work

In the current scenario, our proposed integrated trust mechanism validated with available resources but proposed mechanism is aware of requirement of scientific programming techniques. The proposed trust mechanism requires more improvement and we could involve more factors in calculation of trust opinions which can improve buyer and sellers trust relationship. In the future semantic web technologies could be used to verify our proposed trust mechanism in multi agent environment, obviously little change in design will be require. We will look forward to implement our proposed system with large amount of data in auction systems. The performance of trust opinions requires more resources. There is a need to improve the techniques that are adopted in proposed integrated trust mechanism this may be explored in near future. The experiment was executed with students in University environment; we would look forward to evaluate proposed integrated trust mechanism in industry. The proposed integrated trust mechanism also needs to be validated in auction system industry.

The proposed trust mechanism could preferably use in cloud computing during the exchange of documents/services. Parties who are exchanging their documents or services may take trust factors into account to have trustworthy relations. There are some trust models which may propose for future:

- Trusted medical applications
- Trusted storage services
- Trusted email clients
- Trusted innovative applications

In addition to improve proposed trust mechanism, larger scale opinions and suggestions of industry professionals are welcome to further mature it for industrial use.

# LIST OF FIGURES

# LIST OF TABLES

# APPENDIX A

**Specific Information Related To Research Article**

The specific information about selected research article was documented, are listed below.

**Environment/type of study:**

- Industrial
- Academia
- Consultant report
- Licentiate thesis
- Research Methodology adopted:
- Experiment
- Case Study
- Survey
- Interviews

**Participants of study:**

- Researchers
- Industry professionals
- Students
- Total number of participants

**Relevant area of study:**

- Reputation based mechanism
- Policy based mechanism
- Weaknesses of reputation based trust mechanism
- Strengths of reputation based trust mechanism
- Strengths of policy based trust mechanism
- Weaknesses of policy based trust mechanism
- Comparison of both trust mechanisms

**Results found through each selected electronic database**

Table 10 Results found from each selected database

| Sr. No | Name of Database | Total number of results found | % of results found |
|---|---|---|---|
| 1 | IEEE Explorer | 564 | 23% |
| 2 | ACM Digital Library | 594 | 24% |
| 3 | Inspec (www.iee.org/Publish/INSPEC/) | 885 | 35% |
| 4 | ISI (Online search engine database) | 153 | 6% |
| 5 | EI Compendex (www.engineeringvillage2.com) | 301 | 12% |
| Total | | 2497 | 100% |

**Selected list of Journals, Conferences and Books**

Table 11 Selected list of journals, conferences and books

| JOURNALS |
|---|
| Trust concerns in the Semantic Web |
| Trust negotiation |
| Decentralization and referral trust |
| Computational and online trust models |
| Trust metrics in a web of trust |
| Computational and online trust models |
| Trust in P2P networks and grids |
| Application-specific reputation |
| Filtering information based on trust |
| CONFERENCES |
| International Semantic Web Conference, 2004 |
| IEEE/WIC International Conference on Web Intelligence, 2003 |
| IEEE International Conference on E-Commerce Technology, 2004 |
| International joint conference on Autonomous agents and multi agent systems, 2002 |
| European Conference on Artificial Intelligence, 2004 |
| International Conference on Information and Knowledge Management, 2001 |
| BOOKS |
| A Semantic Web primer, Grigoris Antonius and Frank Van Harmelen |
| Semantic web and semantic web services, Liyang Yu |
| Introduction to Ecommerce, Jeffrey F. Rayport |
| Ontology-Based Policy Specification and Management, O. Daniel, W. Marianne and C. Zhang |
| Semantic Web Services, Processes and Applications, Jorge Cardoso |
| Implementing and Managing E-security, Andrew Nashi and Celia Joseph |

**Selected Research Papers**

Table 12 Selected articles list

| No | Reference No | Publication Year | Selected Research Papers |
|---|---|---|---|
| 1. | 2 | 2007 | A Survey of Trust in Computer Science and Semantic Web |
| 2 | 44 | 2007 | A Trust Based Methodology for Web Service Selection |
| 3 | 41 | 2004 | Can eCRM and Trust improve eC customer base? |
| 4 | 38 | 2005 | Ontology-Based Policy Specification and Management |
| 5 | 43 | 2006 | Semantic Web Policies – A Discussion of Requirements and Research Issues |
| 6 | 40 | 2008 | A Framework for Agent-Based Trust Management in Online Auctions |
| 7 | 42 | 2004 | Using Context- and Content-Based Trust Policies on the Semantic Web |
| 8 | 64 | 2009 | Trust Management in Opportunistic Networks: A Semantic Web Approach |
| 9 | 39 | 2004 | No Registration Needed: How to Use Declarative Policies and Negotiation to Access Sensitive Resources on the Semantic Web |
| 10 | 31 | 2004 | Accuracy of metrics for inferring trust and reputation |
| 11 | 47 | 2009 | Challenges for Robust Trust and Reputation Systems |
| 12 | 45 | 2002 | The value of reputation on eBay: A controlled experiment |
| 13 | 66 | 2004 | Peer Trust: Supporting Reputation-Based Trust for Peer-to-Peer Electronic Communities |
| 14 | 63 | 2008 | Online reputation systems: Design and strategic practices |
| 15 | 68 | 2005 | Reputation Mechanisms |
| 16 | 65 | 2007 | A survey of trust and reputation systems for online service provision |
| 17 | 27 | 2002 | Reputation and social network analysis in multi-agent Systems |
| 18 | 48 | 2004 | How Effective Are Electronic Reputation Mechanisms? An Experimental Investigation |
| 19 | 62 | 2005 | An Integration of Reputation-based and Policy-based Trust Management |
| 20 | 10 | 2009 | An Integrated approach for Trust Management in Semantic Web |
| 21 | 68 | 2004 | Trust Strategies for the Semantic Web |

# APPENDIX B

## Interview Questionnaires

Questionnaire related to benefits of Policy based approach.

- Do you think that Policy based is an easy approach rather than reputation based?
- Do you think that, only policy of a company can improve the security and privacy of the system?
- What do you think disclosing personal information could be a threat, if yes then how can we overcome on this threat?
- Do policy based approach is helpful to increase the market share or business?
- What do you think; any kind of policy assurance from company/third party would be helpful for buyers to make tractions?
- Services of a trusted third party may be used for verification of certificates?
- Do you think policy based approach can improve trust between buyer and seller especially in auction systems?

Questionnaire related to problems of Policy based approach.

- Have you come across with issues ever while registering policy in any auction system?
- What do you think real-world polices are complex in implementations?
- In situation of disclosing your bank account or credit card information in auction system, what could help you to have trust on system?
- New seller without reputation but with strong policy approvals can encourages the buyer to make transactions with that particular seller?
- What do you think a mixed approach, policy and reputation can improve the trust of customer in auction systems?
- Please list down any benefits related to Policy based approach which you have experienced?
- Please mention about the drawbacks associated with mixed approach (policy/reputation), if any?

Questionnaire related to benefits of Reputation based approach.

- Do you think that reputation/rating based is an easy approach rather than policy based?
- Do you think current reputation/rating system is suitable for auction systems (buyer/seller seniors)?
- Do you think the probability of sale may change with the good/bad reputation?
- Do you think reputations/rating system can make an effect on strong partnerships?
- Do you think less rating can discourage people to join the market place?
- Do you think pre designed reputation systems are sufficient for auction systems?
- Do you think rating system is an effective approach for making trust on buyer and seller?
- Do you think your customers are satisfied with current reputation systems?

Questionnaire related to problems of Reputation based approach.

- What factors you think of which can make Reputation mechanism more beneficial?
- Do you think current reputation system is suitable for improve trust of customer on seller in auction systems?
- In what scenarios do you think Reputation is not preferred on Policy based approaches?
- Have you experienced any issues related to Reputation based approach?
- Available rating systems in the market are too positive where negative feedback rarely affects the overall trust of customers on seller?
- Do you think an agent can rate someone multiple times, have you experienced anything like that?
- Buyer or seller can be in shape of re-identity to join again the community; do you have any past experience with these kinds of situations?
- What do you think how new agent can have trust between community of auction systems?

Suggestions

If you have any suggestions which can improve customer's trust on sellers related to auction systems.

# Transcribed Interviews

# Interview 1

**Interviewee**

Name: Asad Masood Khattak
Email address: asadmasood@gmail.com
Contact information: Ubiquitous Computing Lab
Dept. of Computer Engineering
Kyung Hee University, Korea.
Date: November 23, 2010
Start and End time: 10:00pm to 10:45pm

The contacted person for interview was Mr. Asad Masood Khattak, currently working as a researcher in kyung Hee University, Korea. The research area of Mr. Asad is Semantic web and he has more than eight publications till now. The interviewee was an extremely valuable resource because of his experience with semantics.

**Transcribed interview**

In the start we have discussed about the strengths and problems associated with policy based trust mechanism. He told us that policy based trust mechanism is best choice to improve the security and privacy according to company needs. According to him, disclosing personal information is a threat in policy based trust mechanism and suggesting that some encryption mechanism can be used by sharing the description keys. He says, that policy based trust mechanism is a simple approach and help to increase the market shares in case of specific groups. According to him, policy based trust mechanism improve trust between buyer and seller in auction system. He told us that services of a trusted third party may be used for the verification of certificates but policy assurance from third party would not be helpful for buyers to make transactions. According to him, real world polices are not too complex to implement and suggest, it should be decomposed first and can be easily implemented. According to him, strong polices of new sellers encouraged buyers to make transactions with that particular seller.

According to him, rating system is an effective approach for making trust decisions. Good and bad reputation changing the probability of sale, and it is an efficient way to encourage or discourage a seller on the basis of the services he provides. He told us, that strong partnership could be established with high reputation more easily. According to him, the use of multiple identities is the main problem in reputation systems and it should be need to handle this approximately. He told us, that the new user maybe trustworthy if he has good business with any of the existing community member and that he can verify his reputation or new user can bring inherited reputation from the community that he is migrating from. He told us that reputation can't be proffered on policy based mechanism in the scenario, where you have a predefined hard and fast policies which must meet before someone use your services.

According to him, an integrated mechanism of both policy and reputation based trust mechanisms will be beneficial. He encouraged a mixed mechanism that it will handle the problems associated with each of them, e.g., policy is that no dealing with customer if there is one prior problem with that customer during any past dealing but reputation suggest establishing a new deal with the customer.

# Interview 2

**Interviewee**

Name: Adil Farid
Email address: uomian2004@yahoo.com
Contact information: Dept. of Software Engineering
University of L'Aquila, Italy.
Date: November 24, 2010
Start and End time: 08:00pm to 08:35pm

The contacted person for interview was Mr. Adil Farid, studying in University of L'Aquila, Italy. He worked for 3 years as a Web Developer at Dynamism IT Solutions Peshawar Pakistan. Currently he is doing his research in semantic web. Because of his experience and current research he has a sound knowledge and experience of different trust mechanisms being followed in the industry.

**Transcribed interview**

In the start we have discussed about the policy based trust mechanism. According to him, Policy plays an important role in security, privacy and in assigning tasks to the users. He told us that policy based trust mechanism is an easy approach for establishing trust. According to him, disclosing personal information would not be a threat in such a mechanism because a user can only share their information when he assures that it will not be disclosed. He says that specific long term polices increase market shares and using trusted third party services for the verification of credentials is a good choice to increase the trust of the buyer and giving him confidence to make transactions. According to him real-world polices are often hard to implement.

He says that Policies do matter in security and privacy of the system, but as part of the nature, people usually trust in the most experienced ones. A well rated site is preferable then a site with good policies. He says, that good reputation always matters, people prefer to purchase on the sites with good reputation even if they have to pay little extra. High reputation companies often make strong partnerships with each other. He says that, reputation trust mechanism is good to encourage or discourage sellers on the basis of the services they provide. Reputation based trust mechanism is trustable because of the third party involvement as well. He told us that reputation based trust mechanisms still needs some improvements, because still people are reluctant of purchasing online. Using multiple identities is the main problem in reputation based trust mechanism, which must be handled. He says that a mixed mechanism of both policy and reputation based trust mechanisms will be more trustable and reliable. According to him, any mixed approach is always welcomed. A good reputation site with strong policies will be always trustable.

# Interview 3

**Interviewee**

Name: Arif Ur Rahman
Email address: badwanpk@hotmail.com
Contact information: Researcher at INESC Porto
Faculty of Engineering
University of Porto, Portugal
Date: November 26, 2010
Start and End time: 11:00pm to 11:40pm

The contacted person for interview was Mr. Arif Ur Rehman, working as a researcher at INESC Porto, Portugal. He has more than three years experience in client server applications and currently working as a researcher in the same area. Because of his experience and current research he has sound knowledge of different trust mechanisms being followed in the industry.

**Transcribed interview**

We have discussed in detail about both policy and reputation based trust mechanisms. According to him, as specific resource or information can be allowed after the verification of certain credentials that an entity provides in policy based trust mechanism, so it is more efficient for the improvements of security in an organization. He told us that direct trust establishment between seller and buyer would be more efficient instead of using services of a third party in policy based trust mechanism. According to him, different trust levels can be defined on the basis of information that an entity provide in policy based trust mechanism. An entity will be more trustable if he provides some strong information like personal number than an entity who provides name, email address etc. A new seller would be able to attract more customers with the implementation of good/easy polices.

According to him, reputation based trust mechanism as more efficient approach for making trust decisions on the basis of trustable third parties. The probability of sale must be change with these ratings, where a seller having high rating will be able to earn more profit then a seller having low reputation. Sellers having high reputations will be able to establish partnerships more easily with other high rating sellers. According to him, most of the reputation systems in the market are too positive from seller's point of views and the use of multiple identities is the second threat in such reputation based mechanisms. Sellers and buyers having bad reputation often enter the market with different identity which must be handled. He told us that a new entity maybe be trustable in such a reputation based mechanism if he come with some predefine ratings.

He says that every new approach is welcomed and it will be better to have a mixed mechanism of both policy and reputation based trust mechanism. He suggests the use of trust levels in an integrated trust mechanism instead of the verification of credentials from third parties. Second he suggests the implementation of strong polices with whom a user will not be able to use multiple identities. Thus an integrated trust mechanism will be more efficient either from policy or reputation based trust mechanism.

# Interview 4

**Interviewee**

Name: Muhammad Sami Ullah
Email address: samiforlisten@yahoo.com
Contact information: Lecturer in Department of Computer Science,
University of Punjab, Lahore, Pakistan
Date: November 26, 2010
Start and End time: 11:00am to 11:45am

The contacted person for interview was Mr. Muhammad Sami Ullah, lecturer in the department of computer science, University of Punjab, Lahore, Pakistan. He has been worked on many projects i.e. OWL API USAGE, DLD Ontology, LAN Based Client-server Model for Chatting, Shooting Game etc and has four publications in semantic web. The interviewee was an extremely valuable resource because of his experience and research work in semantic web.

**Transcribed interview**

In the start of the interview, we have discussed about the policy based trust mechanism. He told us, that policy based trust mechanism are secure because only trustworthy user were allowed to specific information and for making transactions in this mechanism. New customer will be considered trustworthy because of the verification of their credentials, as well only trustworthy consumers will be encouraged to join the market. With the implementation of strong polices same agent will not be able to enter the market with multiple identities. According to him in policy based trust mechanism, most times information required in pre-registration phase are not specifically relevant to the resources he/she wants to access.

According to him reputation is flexible and simple approach for maintaining trust. He says that reputation based trust approach is more efficient, in which people trust or distrust on a specific seller or item because of their ratings. Feedbacks provide valuable information in such a mechanism, helps buyers in decision making as well sellers to increase/decrease price of an item, change picture, improve quality etc. He told us that unfair rating and the use of multiple identities are the main problems in reputation based mechanism. New sellers having no reputation at all face problem in the start in such a mechanism. He says that new sellers maybe trustworthy if he has good business with any of the existing community member and that he can verify his reputation or new user can bring inherited reputation from the community that he is migrating from. The problem of multiple identities maybe solved with the implementation of strong polices in reputation based mechanism. An integration of both policy and reputation based trust mechanisms is a good idea. The drawbacks associated with each policy and reputation based trust mechanism can be minimized with an intelligent integration. According to him, such an integrated trust mechanism where multiple accounts are not possible will be more beneficial for organizations using reputation based trust mechanism. The involvement of trusted third parties will be another positive aspect in an integrated trust mechanism with respect to policy based trust mechanism.

# APPENDIX C

**Questionnaire for Experiment Evaluation**

Table 13 Questionnaire for experiment evaluation

| No | Questionnaire for Experiment Evaluation | Strongly Agree | Agree | Less Agree | Disagree | Strongly Disagree |
|---|---|---|---|---|---|---|
| 1 | The policy of system supports new seller? | | | | | |
| 2 | I found to have multiple accounts on system is impossible | | | | | |
| 3 | I found trust opinions/information while browsing products is convenient | | | | | |
| 4 | The trust values shown with registration are not different than without registration | | | | | |
| 5 | System supports cost and product category dynamically to have more trust on seller? | | | | | |
| 6 | The cost of product effects reputation, low cost increases low and high cost increases high rate in reputation calculation | | | | | |
| 7 | My general intention to make deal with new seller is high/changed | | | | | |
| 8 | To what extent you can make a deal with new seller? | | | | | |
| 9 | The system have a simple mechanism to remove bad image as a bad rating | | | | | |
| 10 | I would recommend same proposed system for other auction systems | | | | | |
| 11 | Overall, performance of the application was satisfactory | | | | | |

**Frequency Tables**

Table 14 Scale distribution

| Strongly Agree | Agree | Less Agree | Disagree | Strongly Disagree |
|---|---|---|---|---|
| 5 | 4 | 3 | 2 | 1 |

N = Frequency, total number of occurrences for each degree on scale.
% = Relative frequency, dividing each frequency by total no of samples.

Table 15 Registration policy

| Registration Policy | Proposed System | | Tradera | | eBay | |
|---|---|---|---|---|---|---|
| | N | % | N | % | N | % |
| Strongly Agree | 8 | 20 | 6 | 15 | 0 | 0 |
| Agree | 28 | 70 | 25 | 63 | 6 | 15 |
| Less Agree | 4 | 10 | 9 | 22 | 24 | 60 |
| Disagree | 0 | 0 | 0 | 0 | 10 | 25 |
| Strongly Disagree | 0 | 0 | 0 | 0 | 0 | 0 |
| Total | 40 | 100 | 40 | 100 | 40 | 100 |

Table 16 Reputation calculation

| Reputation Calculation | Proposed System | | Tradera | | eBay | |
|---|---|---|---|---|---|---|
| | N | % | N | % | N | % |
| Strongly Agree | 7 | 17 | 0 | 0 | 0 | 0 |
| Agree | 28 | 70 | 19 | 48 | 16 | 40 |
| Less Agree | 5 | 13 | 16 | 40 | 19 | 48 |
| Disagree | 0 | 0 | 5 | 12 | 5 | 12 |
| Strongly Disagree | 0 | 0 | 0 | 0 | 0 | 0 |
| Total | 40 | 100 | 40 | 100 | 40 | 100 |

Table 17 Effect of cost

| Cost Effects | Proposed System | | Tradera | | eBay | |
|---|---|---|---|---|---|---|
| | N | % | N | % | N | % |
| Strongly Agree | 18 | 45 | 0 | 0 | 0 | 0 |
| Agree | 22 | 55 | 0 | 0 | 0 | 0 |
| Less Agree | 0 | 0 | 5 | 12 | 5 | 12 |
| Disagree | 0 | 0 | 22 | 55 | 22 | 55 |
| Strongly Disagree | 0 | 0 | 13 | 33 | 13 | 33 |
| Total | 40 | 100 | 40 | 100 | 40 | 100 |

Table 18 Support for new seller

| New Seller | Proposed System | | Tradera | | eBay | |
|---|---|---|---|---|---|---|
| | N | % | N | % | N | % |
| Strongly Agree | 13 | 33 | 0 | 0 | 0 | 0 |
| Agree | 21 | 53 | 4 | 10 | 0 | 0 |
| Less Agree | 6 | 15 | 10 | 25 | 10 | 25 |
| Disagree | 0 | 0 | 17 | 43 | 21 | 53 |
| Strongly Disagree | 0 | 0 | 9 | 22 | 9 | 22 |
| Total | 40 | 100 | 40 | 100 | 40 | 100 |

Table 19 Bad image (negative rating)

| Bad Image | Proposed System | | Tradera | | eBay | |
|---|---|---|---|---|---|---|
| | N | % | N | % | N | % |
| Strongly Agree | 2 | 10 | 0 | 0 | 0 | 0 |
| Agree | 12 | 60 | 3 | 15 | 3 | 15 |
| Less Agree | 6 | 30 | 8 | 40 | 15 | 75 |
| Disagree | 0 | 0 | 9 | 45 | 2 | 10 |
| Strongly Disagree | 0 | 0 | 0 | 0 | 0 | 0 |
| Total | 20 | 100 | 20 | 100 | 20 | 100 |

Table 20 Customer satisfaction

| Customer Satisfaction | Proposed System | | Tradera | | eBay | |
|---|---|---|---|---|---|---|
| | N | % | N | % | N | % |
| Strongly Agree | 11 | 28 | 5 | 13 | 4 | 10 |
| Agree | 26 | 65 | 22 | 55 | 19 | 48 |
| Less Agree | 3 | 7 | 13 | 32 | 14 | 35 |
| Disagree | 0 | 0 | 0 | 0 | 3 | 7 |
| Strongly Disagree | 0 | 0 | 0 | 0 | 0 | 0 |
| Total | 40 | 100 | 40 | 100 | 40 | 100 |

# APPENDIX D

## Products List with Score

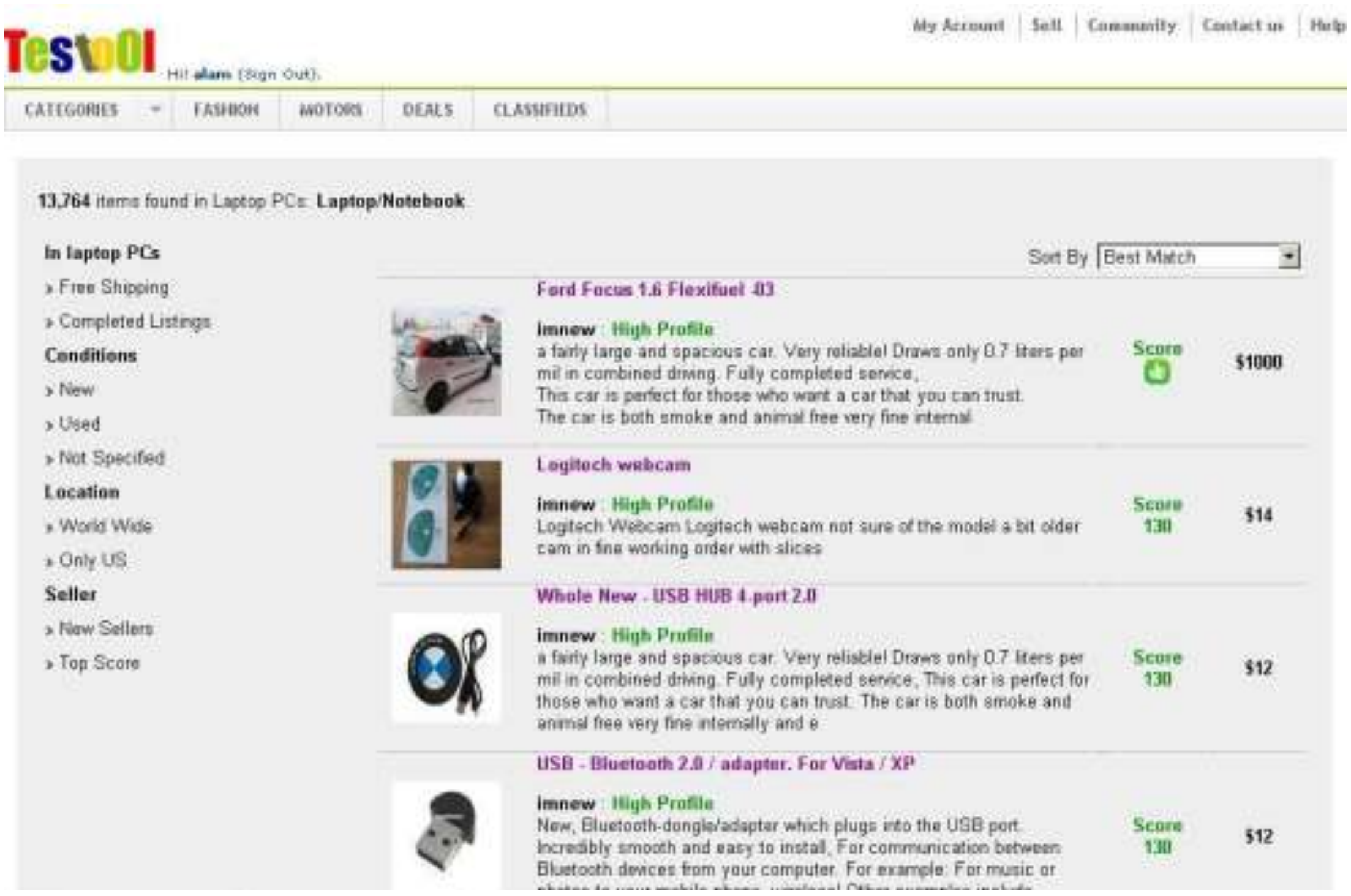

## Product Detail with trust opinions

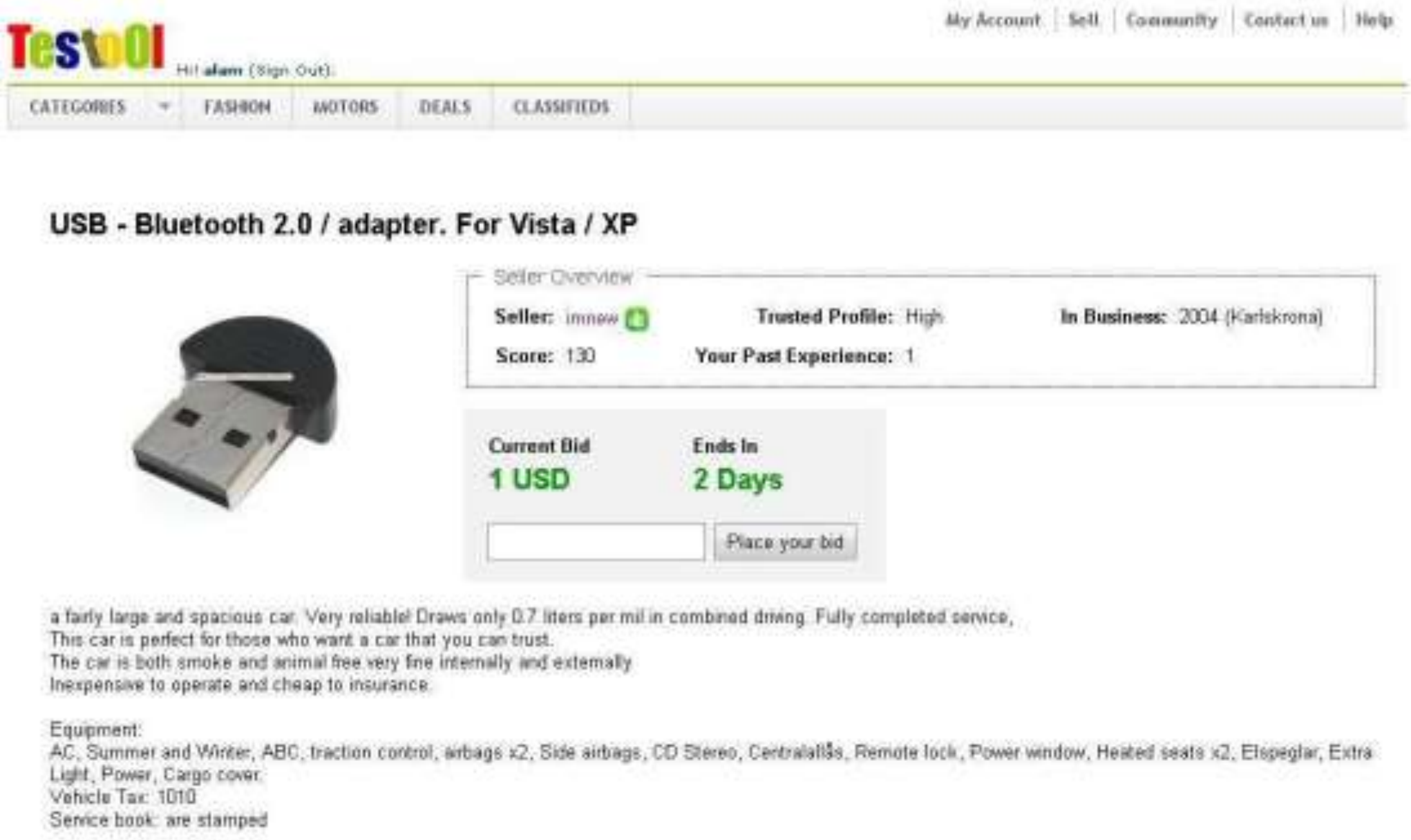

## Product details with different trust opinions and different scope of products

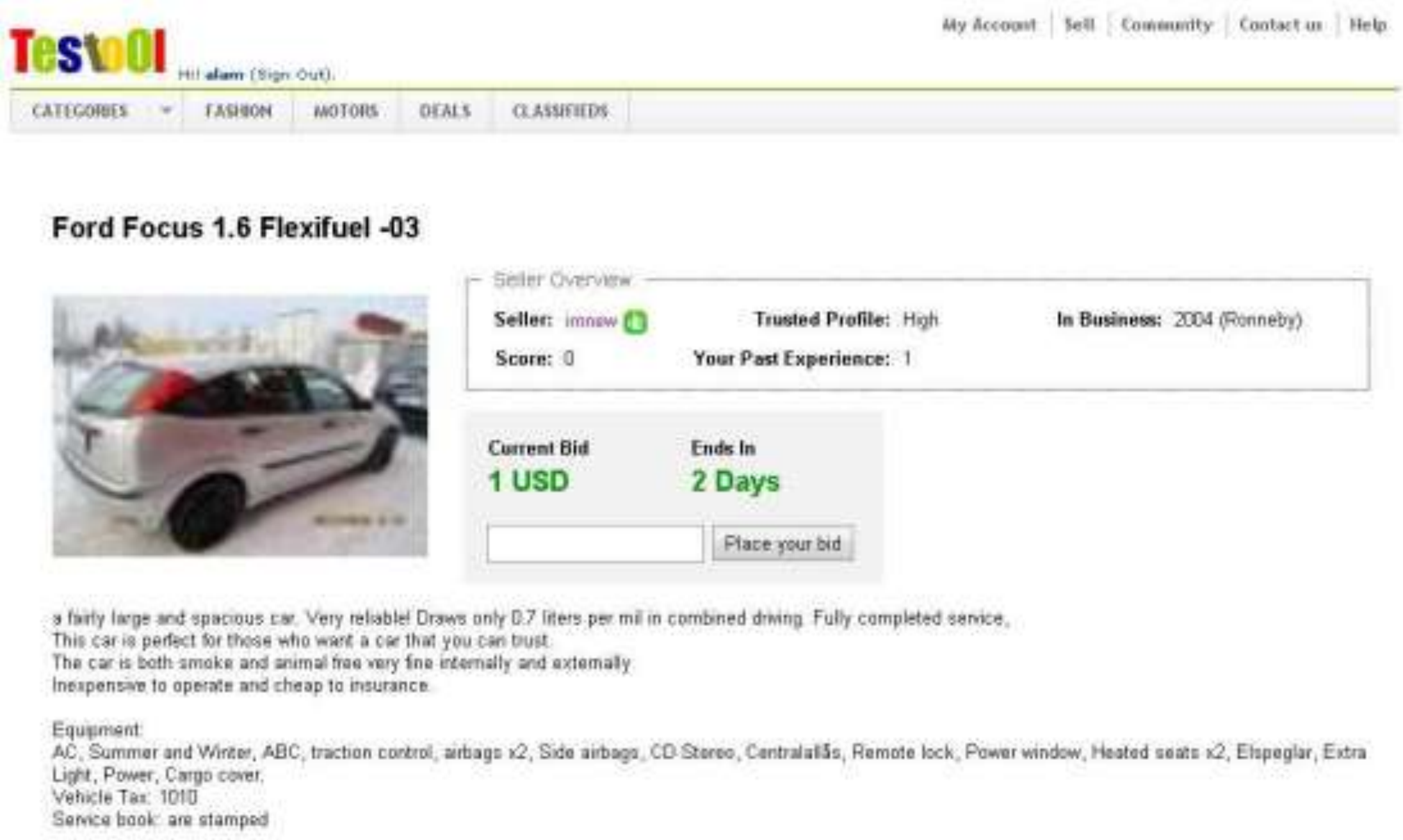

## Policy based trust registration mechanism

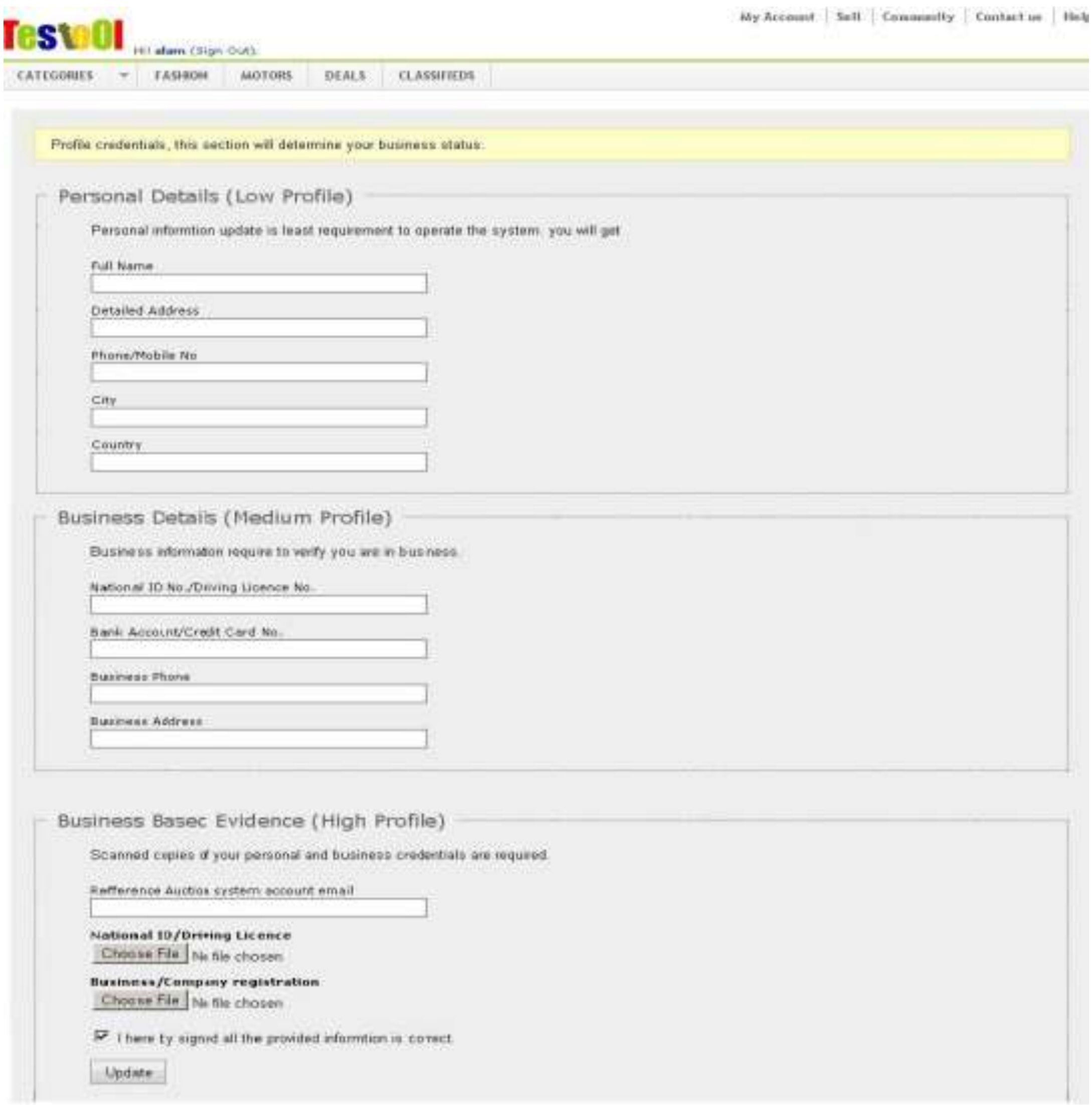